\documentclass[useAMS,usenatbib]{mn2e}

\usepackage{lscape}
\usepackage{longtable}

\usepackage{graphicx}

\title{Galaxy clusters and groups in the ALHAMBRA Survey}
\author[B. Ascaso et al.]{B. Ascaso$^{1,2}$\thanks{E-mail:
begona.ascaso@obspm.fr},  
N. Ben\'itez$^{2}$, A. Fern\'andez-Soto$^{3,4}$, P. Arnalte-Mur$^{5,6}$, 
 \newauthor C. L\'opez-Sanjuan$^{7}$,  A. Molino$^{8,2}$,  W. Schoenell$^{2}$,   Y. Jim\'enez-Teja$^{9,2}$,   A. I. Merson$^{10}$, 
 \newauthor    M.~Huertas-Company$^{1,11}$,  L.~A.~D{\'{\i}}az-Garc{\'{\i}}a$^{7}$, V. J. Mart\'inez$^{4,12}$, A. J. Cenarro$^{7}$,     
\newauthor R. Dupke$^{8,13,14}$, I. M\'arquez$^{2}$,  J. Masegosa$^{2}$, L. Nieves-Seoane$^{3,5}$, M. Povic$^{2}$,    
\newauthor  J. Varela$^{7}$, K. Viironen$^{7}$,  J.A.L. Aguerri$^{15,16}$, A. Del Olmo$^{2}$, M. Moles$^{7,2}$, J. Perea$^{2}$,   
\newauthor E.~Alfaro$^{2}$, T.~Aparicio-Villegas$^{8,2}$, T.~Broadhurst$^{17,18}$, J.~Cabrera-Ca\~no$^{19}$, 
\newauthor F.~J.~Castander$^{20}$, J.~Cepa$^{15,16}$, M.~Cervi\~no$^{2,15,16}$, R.~M.~Gonz\'alez~Delgado$^{2}$, 
\newauthor D.~Crist\'obal-Hornillos$^{7}$,  L.~Hurtado-Gil$^{3,5}$, C.~Husillos$^{2}$, L.~Infante$^{21}$, F.~Prada$^{2,22,23}$,   \newauthor J.~M.~Quintana$^{2}$}

\begin{document}

\date{Accepted . Received }


\maketitle

\label{firstpage}

\begin{abstract}

We present a catalogue of 348 galaxy clusters and groups with $0.2<z<1.2$ selected in the 2.78 $deg^2$ ALHAMBRA Survey. The high precision of our photometric redshifts, close to $1\%$, and the wide spread of the seven ALHAMBRA pointings ensure that this catalogue has better mass sensitivity and is less affected by cosmic variance than comparable samples. 

The detection has been carried out with the Bayesian Cluster Finder (BCF), whose performance has been checked in ALHAMBRA-like light-cone mock catalogues. Great care has been taken to ensure that the observable properties of the mocks photometry accurately correspond to those of real catalogues. From our simulations, we expect to detect galaxy clusters and groups with both $70\%$ completeness and purity down to dark matter halo masses of $M_h\sim3\times10^{13}\rm M_{\odot}$ for $z<0.85$. Cluster redshifts are expected to be recovered with $\sim0.6\%$ precision for $z<1$. We also expect to measure cluster masses with $\sigma_{M_h|M^*_{CL}}\sim0.25-0.35\, dex$ precision down to $\sim3\times10^{13}\rm M_{\odot}$, masses which are $50\%$ smaller than those reached by similar work. 

We have compared these detections with previous optical, spectroscopic and X-rays work, finding an excellent agreement with the rates reported from the simulations. We have also explored the overall properties of these detections such as the presence of a colour-magnitude relation, the evolution of the photometric blue fraction and the clustering of these sources in the different ALHAMBRA fields. Despite the small numbers, we observe tentative evidence that, for a fixed stellar mass, the environment is playing a crucial role at lower redshifts (z$<$0.5).

\end{abstract}


\section{Introduction}

Galaxy clusters are the largest objects gravitationally bound in the universe. According to the standard model of cosmic structure formation, they appeared as a result of the initial perturbations in the mass power spectrum at a typical comoving scale of $\simÊ10h^{-1}$ Mpc. At larger scales, the universe is mainly dominated by gravity with the gas dynamics a minor contributor. However, at smaller scales, the complexity of the astrophysical processes, particularly related with formation and evolution of galaxies, produces changes in the observational properties of the structures. 

The first catalogues of galaxy clusters came by the hand of  Abell \citep{abell58} and Zwicky \citep{zwicky61} in the early sixties, together with posterior improved extensions expanded  to the southern sky  \citep{abell89}. These catalogues, in addition to suffering from large projection effects due to the absence of multi-band photometry or spectroscopy, had a complicated selection function since they were based on visual inspections  (\citealt{katgert96} and references herein). 

Subsequent cluster and group catalogues have been built from systematic searches of galaxy clusters in different wavelengths or with different techniques: (1) searches in optical data (for a review of the different methodologies in the optical, see \citealt{gal06,ascaso13}); (2) searches with X-ray data, (see \citealt{rosati02,burenin07} and references herein); (3) searches using radio sources (e.g. \citealt{galametz09,chiaberge10,castignani14,blanton14}) (4) searches using the Sunyaev-Zel'dovich (SZ) signature in cosmic microwave background maps (e.g. \citealt{bartlett04,staniszewski09}); (5) searches using the Weak-Lensing (WL) effect (see  \citealt{wittman06} for a detailed explanation and references herein); (6) spectroscopic searches (see, for instance \citealt{knobel09,knobel12} and references arising from them). All these different techniques provide well-characterized selection functions, completeness and purity rates.

Since the beginning of the XXI century, we have been witnesses of the discovery of even more extreme clusters. For the first time, high redshift ($z>1$) structures in the universe using optical/IR data, X-rays, radio or the SZ effect (\citealt{galametz09,fassbender11,planck11,jee11,brodwin13,castignani14}, to name a few) have been found. The finding of these clusters, usually very massive (several times $10^{14}\rm M_{\odot}$), has already impulsed a change in the main theories of cluster formation and galaxy evolution. For instance, the discovery of very massive clusters such as XMMU J2235.3 - 2557 at z $\sim$1.4 \citep{jee09} or `El Gordo' at $\sim$0.87  \citep{menanteau12}, has challenged the main cosmological theories \citep{hoyle12}. The first proto-clusters at redshift $>$ 2 have also set upper limits in the time scale of structure formation (e.g. \citealt{capak11,chiang14}). Complementarily, the discovery of  post-merger clusters (e.g. the `Bullet Cluster', \citealt{bradac08}, the `Musket Ball cluster', \citealt{dawson12}) has shown that clusters are far from being static entities, being able to merge and dramatically change their properties. 

The observational strategies followed to construct the present cluster surveys have favored the detection of the most massive and luminous clusters due to flux limits and resolution effects. Only spectroscopic searches  \citep{knobel09,knobel12} and recent X-ray surveys  \citep{finoguenov07}, have provided complete group catalogues down to low masses ($\sim 10^{13}\rm M_{\odot}$). Unfortunately, the observational cost of these surveys is very high and only small areas have been sampled.

In the last few decades, several multiple medium-bands surveys have been developed: the Classifying Objects by Medium-Band Observations in 17 Filters survey (COMBO-17, \citealt{wolf03}); the Cosmic Evolution Survey (COSMOS, \citealt{scoville07a}); the Advanced Large, Homogeneous Area Medium Band Redshift Astronomical survey (ALHAMBRA, \citealt{moles08}); or the Survey for High-z Absorption Red and Dead Sources (SHARDS, \citealt{perez-gonzalez13}), among others. One of the main benefits of having a large number of medium-band filters covering the whole optical spectrum at least is the fact that the photometric redshift resolution becomes comparable to that of a spectroscopic survey (see \citealt{molino14} for a review) allowing us to sample low-massive structures in the universe within larger areas than spectroscopic samples.

In this work, we have used the ALHAMBRA survey to perform a systematic search of galaxy clusters and groups using the Bayesian Cluster Finder (BCF, \citealt{ascaso12}). The ALHAMBRA survey consists of four square degrees imaged in 20 optical narrow bands and three broad-band IR bands. For technical reasons, only three degrees were observed and calibrated. In addition, part of the images were masked to take into account image artifacts and saturated stars, resulting into a final usable area of 2.78 square degrees. The survey is complete down to $F814W<24.5$, where $F814W$ is a synthetic combined band \citep{molino14} equivalent to the band with the same name at the Hubble Space Telescope (HST). The overall photometric redshift accuracy obtained for the survey is $\Delta z/(1+z_s)<0.015$ \citep{molino14}, making it comparable to low resolution spectra for each object in the survey. This survey, together with its future `big brother', the Javalambre- Physics of the accelerated universe Astrophysical Survey (J-PAS, \citealt{benitez14}) will be able to set a benchmark in the determination of the cluster mass function in surveys. 

The structure of the paper is as follows. In section \S2, we describe the ALHAMBRA dataset. Section \S3 provides the basic information about the BCF cluster finder. Section \S4 is devoted to the description of the mock catalogues used in this work on one hand and, on the other, to the results regarding cluster detection on them, the description of the selection function for the ALHAMBRA survey and the accuracy in measuring cluster properties such as redshift or mass. In Section \S5, we present the ALHAMBRA optical cluster and group detections. We first compare the detections found in this work with those found by other authors using different datasets and techniques. Then, we explore the main properties of the detections such as the stellar mass distribution, the presence of colour-magnitude relations and the fraction of blue and late-type galaxies in the cluster. Finally, section \S6 provides the main conclusions of the paper and includes a discussion of the main results of this work. Where appropriate, we have used $H_0$=73 km s$^{-1}$ Mpc$^{-1}$, $\Omega_{\rm M}$ =0.25, $\Omega_{\Lambda}$=0.75 throughout this paper in order to match the same cosmology than the mock catalogues utilized in this work.

\section{The ALHAMBRA Survey}

The Advanced Large, Homogeneous Area Medium Band Redshift Astronomical (ALHAMBRA\footnote{http://alhambrasurvey.com/}, \citealt{moles08}) survey is a 20 narrow-band optical and three broad-band NIR (JHK) photometric survey imaged with the wide-field cameras LAICA and OMEGA-2000 respectively, at the Calar Alto Observatory (Spain)\footnote{http://www.caha.es/}.

This survey covers 4 square degrees spread in eight different regions of the sky down to $\sim24.5$ AB magnitude in the synthetic combined $F814W$ band and $\sim20.5$ in the infrared bands. Due to technical issues related with the unavailability of calibration stars in one of the fields (ALH-1) and the lack of two pointings in the ALH-4 and ALH-5 fields, only three degrees were available. Also, additional areas of the images such as saturated stars and image edges with insufficient exposure time to provide accurate photometry have been properly masked, obtaining a final area of 2.78 square degrees \citep{molino14}.

The different fields of the ALHAMBRA survey were chosen strategically in order to overlap with well-known fields, many of them with multiwavelength data available such as the Sloan Digital Sky Survey (SDSS, \citealt{york00,ahn14}), the DEEP2 Galaxy Redshift Survey \citep{newman13} or the COSMOS survey \citep{scoville07a}.

The photometric redshifts obtained for this survey have been calculated with BPZ2.0 (Benitez in prep), an improved version of BPZ \citep{benitez00} capable of also providing stellar masses. The catalogues, including a full range of measured properties of the galaxies, are publicly available\footnote{https://cloud.iaa.csic.es/alhambra/} (see \citealt{molino14} for details). Additionally, a set of mask files including saturated star and spurious effects are also available for the survey. These were generated together with the mock catalogues and their building procedure can be found in \citealt{molino14}.

\section{The Bayesian Cluster Finder}

In this work, we have used the Bayesian Cluster Finder (BCF; \citealt{ascaso12}), a technique developed to detect galaxy clusters and groups in any optical/infrared image dataset. This method is based on a modification of the Matched Filter Technique \citep{postman02} including photometric redshifts and the presence of Bayesian priors. 

In more detail, the BCF initially calculates the probability that there is a cluster centered on each galaxy at a given redshift. In order to calculate this likelihood, we assume that clusters are modeled as a convolution of a particular density, luminosity and photometric redshift profile. We choose to use the Plummer density profile \citep{postman02}, the Schechter luminosity function \citep{schechter76} and the full redshift probability function  (P(z), \citealt{benitez00,molino14}) in case it is available or a Gaussian approximation otherwise. This likelihood probability does not include any pre-assumption about the colours of the cluster. We convolve this likelihood with a prior probability to obtain the final probability. The prior refers to those properties that are not necessarily present in all clusters but can help to discern between different solutions. We choose to model two main features that are present in the majority of the clusters up to redshift $\sim 1.6$ and down to masses $10^{14}\rm M_{\odot}$ at least: the presence of the colour-magnitude relation (CMR; e.g. \citealt{lopez-cruz04,mei06,ascaso08,papovich10}) and the presence of a well-defined brightest cluster galaxy (BCG), following a tight relation between its magnitude and the redshift of the cluster (e.g \citealt{ascaso11,ascaso14b}).

In order to characterize the former prior component, we first created synthetic $g-i$ and $i-z$ colours predictions using a typical elliptical spectrum from the library by  \cite{coleman80} as performed in \cite{ascaso12}. We chose those bands in order to be able to sample the 4000$\AA$ break at low ($z<0.7$) and high ($z>0.7$)-redshift range respectively. We artificially created these bands by calculating the contribution of each of the ALHAMBRA narrow bands to the new synthetic band as performed in \cite{molino14}. We then created a Gaussian filter characterizing the colour for each redshift slice. The width of this Gaussian has been set to 0.5 in order to account for larger dispersions in the RS. As for the latter prior feature, we first measured empirically the K-band Hubble diagram for a complete sample of BCGs extracted from \cite{whiley08}, \cite{stott08} and \cite{collins09} up to redshift $<1$. Then, we performed a colour transformation to our reference band, $F814W$ using the same library of synthetic templates as for the CMR. Similar to the case of the CMR, we created a Gaussian filter characterizing the magnitude of the BCG at each redshift slice. Note that for a real detection, the amplitude of the likelihood is always several orders of magnitude larger than the signal of the prior. As a consequence, the prior information helps to discern between different solutions at different redshift slices without losing preliminary detections by the original likelihood. In other words, the BCF does not specifically rely on the colours of the galaxies to select clusters, therefore it is able to detect any structure over the threshold limit, independent of its colour.

We performed a search in a predefined number of redshift slices. The minimum redshift threshold ($z_{min}$) comes from the angular extent of the survey which is limited by its geometry; and the maximum redshift ($z_{max}$) is estimated from the wavelength coverage and the depth of the survey. The bin width ($z_{bin}$) is fixed to be three times the expected photometric resolution of the survey. For instance, for the ALHAMBRA survey, we have fixed $z_{min}=0.2$, $z_{max}=1.2$ and $z_{bin}=0.05$. Effects of stars masking and edges of the frames are taken into account \citep{molino14}. As in \cite{ascaso12,ascaso14a}, we applied a probability correction proportional to the effective area within 0.5 Mpc of the considered galaxy. This correction accounts for galaxies lying close to the border of the image or saturated stars.

Afterward, clusters are selected as the density peaks of those probability maps and the center is located at the peak of the probability. Finally, if we find a cluster or group detected in different redshift slices (two or more detections separated by less than 0.5 Mpc and one redshift bin difference), we merge them into one. For a more detailed description of the method, we refer the reader to \cite{ascaso12,ascaso14a}.

The output of the algorithm consists of the centered position, a measurement of the redshift and a measurement of the richness of the cluster. We have used the $\Lambda_{\rm CL}$ parameter, defined as
\begin{equation}
\Lambda_{\rm CL}=\frac{\sum_{i=1}^N L_i (R<0.5 Mpc)}{L^*}
\end{equation}
i.e., the sum of the luminosity of the galaxies (in the $F814W$ band) statistically belonging to the cluster divided by the characteristic luminosity $L^{*}$ \citep{postman02}. The galactic population of the cluster is defined as those galaxies lying within a given radius (0.5 Mpc, for the richness calculation) with a given cut in photometric redshift odds (\emph{odds}$>0.5/(1+z)$, in this work) and that the difference with their photometric redshift and the redshift of the cluster is
\begin{equation}
\label{eq:zdiff}
|z-z_{\rm CL}|<z_{\rm bin}/2+\sigma_{\rm NMAD}(1+z_{\rm CL})
\end{equation}
where $z$ is the redshift of the galaxy, $z_{\rm CL}$ is the redshift of the cluster, $z_{\rm bin}$ is the redshift bin used for the detection and $\sigma_{\rm NMAD}$ is the expected photometric redshift accuracy of the survey (0.0125 for the case of ALHAMBRA). The latter parameter, $\sigma_{\rm NMAD}$, is defined as

\begin{equation}
\label{eq:sNMAD}
\sigma_{\rm NMAD}=1.48\times median\bigg(\frac{|d-<d>|}{1+z_s} \bigg)
\end{equation}
and $d=z_{\rm CL}-z_s$ (see, for instance \citealt{molino14}).

In this work, we have introduced a new richness measurement in addition to the $\Lambda_{\rm CL}$ parameter, the cluster total stellar mass, $M^*_{\rm CL}$, 
\begin{equation}
M_{\rm CL}^*=\sum_{i=1}^N M^*_i (R<0.5 Mpc)
\end{equation}
i.e, defined as the sum of all the stellar masses of the galaxies, $M^*$, statistically belonging to the cluster. The stellar masses have been calculated with BPZ2.0 in the same way as \cite{molino14}.
 
So far, the BCF has been applied to two more optical surveys: a wide survey, the CFHTLS-Archive Research Survey (CARS, \citealt{erben09,ascaso12}), and a very deep survey, the Deep Lens Survey (DLS, \citealt{wittman02,ascaso14a}). In this work, we apply the BCF to a high photometric redshift resolution survey, the ALHAMBRA survey.  The comparison between these studies will indicate the benefits and drawbacks of using datasets with different properties for detecting galaxy clusters and groups in the optical.

\section{Simulations}

In this section, we first describe the light-cone mock catalogues that we have used to mimic the ALHAMBRA data (\S4.1). Then, we use the BCF to detect galaxy clusters and groups in those mocks and obtain accurate cluster and group selection functions (\S4.2) and we finally explore the dark-matter halo mass-richness relation in \S4.3.

\subsection{Light-cone mock catalogue}

The light-cone mock catalogue that we have utilized in this analysis has been obtained from \cite{merson13} (see also \citealt{arnalte-mur14}). This mock galaxy catalogue has been built from a semi-analytical model of galaxy formation, applied to the halo merger trees extracted from a cosmological N-body simulation. The semi-analytical model used is the \cite{lagos11} variant of the semi-analytical galaxy formation model  GALFORM \citep{cole00}, which models the star formation and merger history for a galaxy. Among other physical processes, this model includes feedback as a result of SNe, active galactic nuclei (AGN) and photo-ionization of the intergalactic medium. The model predicts the star formation history of the galaxy and therefore its spectral energy distribution (SED). The population of dark matter (DM) haloes for the mock catalogue comes from the Millennium Simulation \citep{springel05}, a $2160^3$ particle N-body simulation of the $\Lambda$ Cold Dark Matter cosmology starting at $z=127$ and models the hierarchical growth to the present day. The halo merger trees are constructed using particle and halo data stored at 64 fixed epoch snapshots spaced logarithmically in expansion factor. The minimum halo resolution is 20 particles, corresponding to $1.72 \times 10^{10} h^{-1} \rm M_{\odot}$. Finally, the light-cone was constructed from this simulation by replicating the simulation box and choosing an orientation. In addition, a flux cut at $F814W<24.5$ AB was applied to reproduce the selection of the ALHAMBRA survey. The final mock catalogue is limited to $z<2$ and it does not include stars. All the details can be found in \cite{merson13}.
 
As performed in \cite{arnalte-mur14}, we created fifty non-overlapping realizations of the ALHAMBRA survey mimicking its geometry from the whole mock catalogue. Besides, we manually included saturated stars and edge effects in the mock catalogues resembling the masks extracted from the ALHAMBRA data. 

Initially, we ran the well-known photometric redshift code, the Bayesian Photometric Redshift (BPZ2.0, \citealt{benitez00}, Benitez in prep). This code, the same used to calculate the ALHAMBRA photo-z, has a library of empirical templates with a very low outlier rate in high quality photometric catalogues ($<1\%$ in the \citealt{ilbert09} COSMOS catalogue, Benitez, private communication; $<2\%$ in the ALHAMBRA data, and most of those seem to be stars or active galactic nuclei (AGN), see \citealt{molino14}).

Since photometric redshifts are exquisitely sensitive to any discrepancy between the template library and actual galaxy photometry, it can be concluded that the rather sparse (11) BPZ2.0 library contains a complete (up to a few $\%$), even if coarse-grained, representation of real galaxy colours for the galaxy populations sampled by the intersection of ALHAMBRA and the spectroscopic redshift catalogues used to measure the outlier rate. Therefore, any systematic mismatches between galaxy colours in mocks and the BPZ2.0 template library signal the presence of non-realistic galaxy types in those mocks. Note than non-realistic does not mean that those types are physically absurd; only that its aggregated frequency nature is significantly below the $1-2\%$ outlier rate we observe in the real world.  

This is exactly what we find when we run BPZ2.0 on the original mock catalogue, obtaining a photometric redshift accuracy of $\Delta z/(1+z) = 0.0319$, a factor of $\sim 3$ worse than in the real data \citep{molino14}. A closer inspection of the original photometry included in the light-cone, evidenced the above mentioned fact that the SED of a significant fraction of the galaxies included in this mock did not match any of the spectra in the BPZ.2.0 library. Therefore, to obtain a galaxy colour distribution which accurately resembles our real ALHAMBRA catalogues, we had to force these objects into realistic colours, assigning then the closest BPZ2.0 SED. This technique, called \texttt{PhotReal}, has already been applied to previous work  \citep{arnalte-mur14,zandivarez14,ascaso15}, will be fully detailed in a future publication (Benitez et al. in prep). 

Once the SED is thus chosen, we generated galaxy fluxes through the ALHAMBRA set of filters and add to them empirically calibrated photometric noise. This noise is a combination of the expected photometric noise from the observed relationship between magnitudes and errors in the ALHAMBRA filters, plus a systematic noise which is approximately constant with magnitude and most likely unavoidable when measuring galaxy colours in multiband photometry. This systematic is empirically calibrated to be $8\%$ for bluer objects and $6\%$ for red galaxies (Benitez, in preparation).  

\begin{figure}
\centering
\includegraphics[clip,angle=90,width=1.05\hsize]{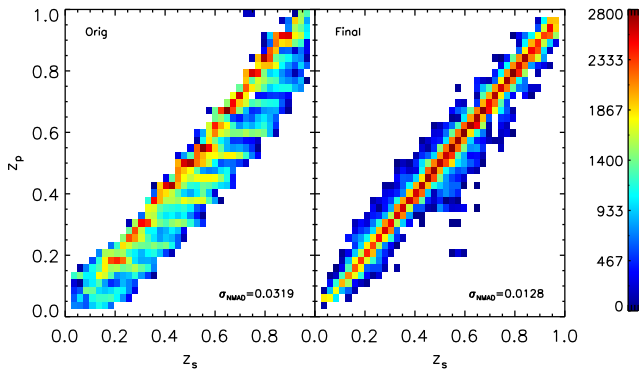} 
\caption{Density plots of the photometric redshift versus spectroscopic redshift for the initial mock catalogue (left panel) and final photometry corrected mock catalogue (right panel). The density scale is shown in the right part of the plot and the overall photometric redshift resolution ($\sigma_{\rm NMAD}$) is quoted (see Eq. \ref{eq:sNMAD}). The photometry correction clearly improves the quality of the photometric redshift and make them comparable to the data.}
\label{fig:photozs}
\end{figure}

Afterwards we run BPZ2.0 on those mock catalogues. In Fig. \ref{fig:photozs}, we display the density maps of the photometric redshift distribution versus spectroscopic redshift for the initial mock catalogue (left panel) and the final mock catalogue (right panel) together with the overall photometric redshift resolution. We notice how, after performing this technique, we increase the resolution by a factor of $\sim3$, reaching values which are very similar to those of real data.  The fact that the real and simulated photometric redshift values are so similar strongly support the fact that the BPZ2.0 library is a faithful representation of the real galaxies.

This procedure of course implicitly introduces an error: there may be real galaxies which are present in the catalogues, but not covered by the BPZ2.0 templates or not adequately present in the spectroscopic redshift catalogues used to measure the outlier rate. Since we are not aware of any such substantial population within the ALHAMBRA depth, we estimate that the procedure followed here may introduce at most a contamination of a few percent, much smaller than that produced by other sources.

As stressed in \cite{molino14} with the ALHAMBRA data, we can additionally increase this resolution by performing a cut in the \emph{odds} parameter. This parameter is defined as the integral of the redshift probability function $P(z)$ around its maximum peak within twice the expected photo-z accuracy for the survey/data which, in the case of the ALHAMBRA data, it was measured to be 0.0125. The \emph{odds} parameter gives us a direct estimation of the quality of the photometric redshift since it measures how concentrated around the `true' photometric redshift value the P(z) is \citep{benitez00,benitez09,molino14}. For instance, performing a cut in \emph{odds} $>0.5$ in the simulation increases the photo-z resolution to $\Delta z/(1+z) = 0.0098$ while a cut in  \emph{odds} $>0.9$ provides a subsample of even higher photometric resolution, $\Delta z/(1+z) = 0.0047$. Of course, a precise understanding of the selection function that the \emph{odds} parameter introduces needs to be quantified. However, this parameter becomes very useful when detecting galaxy clusters for two main reasons: first, it decreases the size of the sample but still leaves `useful' galaxies to detect galaxy clusters and second, we get rid of the field galaxy contamination which have usually lower \emph{odds} parameter values on average. We will come back to this later in the section.

\begin{figure}
\centering
\includegraphics[clip,angle=0,width=1.0\hsize]{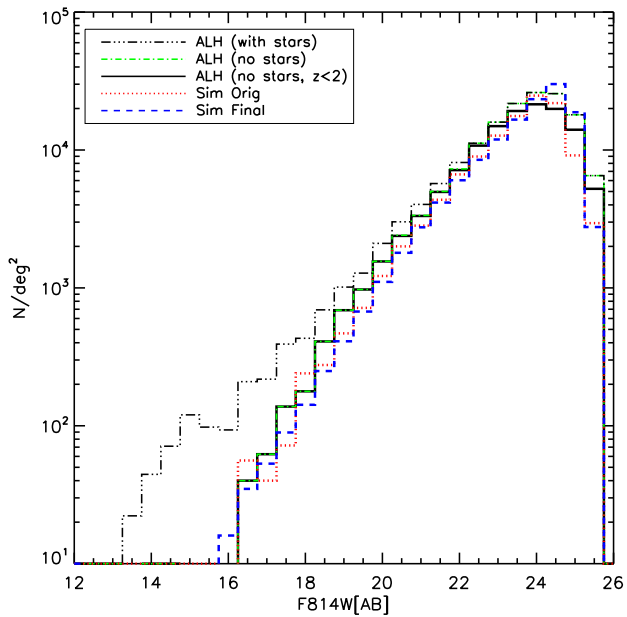} 
\caption{$F814W$ magnitude distribution for the ALHAMBRA data and the mock catalogue data. The dotted-long-dashed black line and the green dotted-short-dashed line refers to the whole ALHAMBRA sample including and excluding the stars, respectively. The black solid line, refers to the ALHAMBRA sample excluding the stars up to redshift 2. The red dotted line and blue dashed line displays the $F814W$ magnitude distribution for the original and final mock catalogues, respectively. We find a very good agreement ($<$8\%) between the $F814W$ distribution for the simulation and the ALHAMBRA data sample up to redshift 2, confirming the accuracy of the mock catalogue.}
\label{fig:magdist}
\end{figure}

Furthermore, we have performed two additional checks for this mock catalogue. The first one has been the comparison of the number counts per square degree between the mock catalogue and the real data as shown in Fig. \ref{fig:magdist}. As we see here, the mock catalogue $F814W$ synthetic magnitude distribution traces almost perfectly the observed one once we removed the effect of stars by making a cut in the stellar flag $<0.7$. The restriction of the ALHAMBRA data to $z<2$ is plotted in order to match the  redshift limit of the mock catalogue. The simulation counts are only $\sim$ 8\% smaller than the real data, a discrepancy that could be attributed to cosmic variance (e.g. \citealt{lopez-sanjuan14}).

\begin{figure}
\centering
\includegraphics[clip,angle=0,width=1.0\hsize]{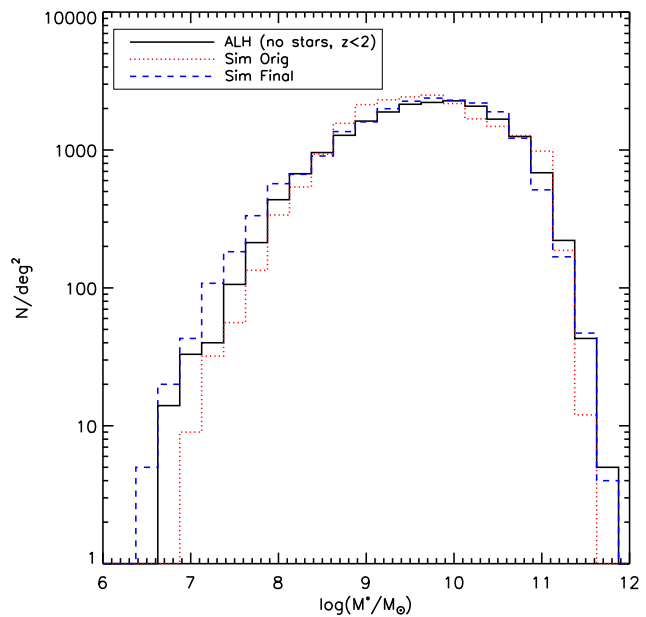} 
\caption{Stellar mass distribution for the ALHAMBRA data and the mock catalogue. The black solid line refers to the whole ALHAMBRA excluding the stars up to redshift 2. The red dotted line alludes to the GALFORM stellar mass distribution in the original mock catalogue and the blue dashed line traces the stellar mass distribution for the post-processed mock catalogue. All the distributions are restricted to the magnitude limit where ALHAMBRA is complete ($F814W<24.5$). We confirm a good agreement between the stellar mass distribution of the ALHAMBRA data and the final mock catalogue ($<$8\%), finding a higher difference ($<17\%$)  between the ALHAMBRA data and the initial mock catalogue.}
\label{fig:massdist}
\end{figure}

The second check performed has been to compare the stellar masses provided by BPZ2.0 in the mock catalogue with the stellar masses in the ALHAMBRA data. In Fig. \ref{fig:massdist}, we display the stellar mass histograms for the ALHAMBRA data, together with the initial simulation and the post-processed simulation.  The stellar masses of the initial simulation refer to those provided by GALFORM. The three distributions have been restricted to $F814W<24.5$, the completeness limit  for the ALHAMBRA survey. We find that the agreement of the observed and post-processed stellar mass distribution is very good ($<$8\%), while the difference is higher for the observed and initial stellar mass distribution ($<$17\%). This latter difference has already been reported in \cite{mitchell13} when comparing the GALFORM stellar masses with stellar masses estimated using SED fitting, and could be due to the dust extinction applied to the model galaxies, which can be substantial for massive galaxies. The disagreement found at the low-mass end of the distribution can be justified as both, the survey and the mock catalogues are flux-limited.

\subsection{Detecting clusters in the mock catalogue}

We have considered as the reference sample in the simulation those haloes more massive than $M_h \ge 10^{13} \rm M_{\odot}$, together with the galaxies associated to each halo and the position of the center, set in one of the galaxies.  Then, we have used the BCF, as described in section \S3, to search for galaxy clusters and groups in the ALHAMBRA mock catalogue restricted to $odds>0.5/(1+z)$. This $odds$ cut has been performed in order to use the best photometric redshift quality galaxies and keep a constant galaxy density as a function of redshift.

We have run the BCF on a test sample using both redshift filters: the full redshift probability function (PDZ) and a simple Gaussian centered in the redshift of the cluster. In this case, the results are very similar ($<$5\% difference). Although a simple Gaussian is not the best approach in photometric surveys  \citep{lopez-sanjuan14,molino14}, the bright galaxies centered in clusters present high S/N ratios, making reasonable the approximation. Therefore, we decided to use the second approach for computational and disk space purposes.

\begin{figure}
\centering
\includegraphics[clip,angle=0,width=1.0\hsize]{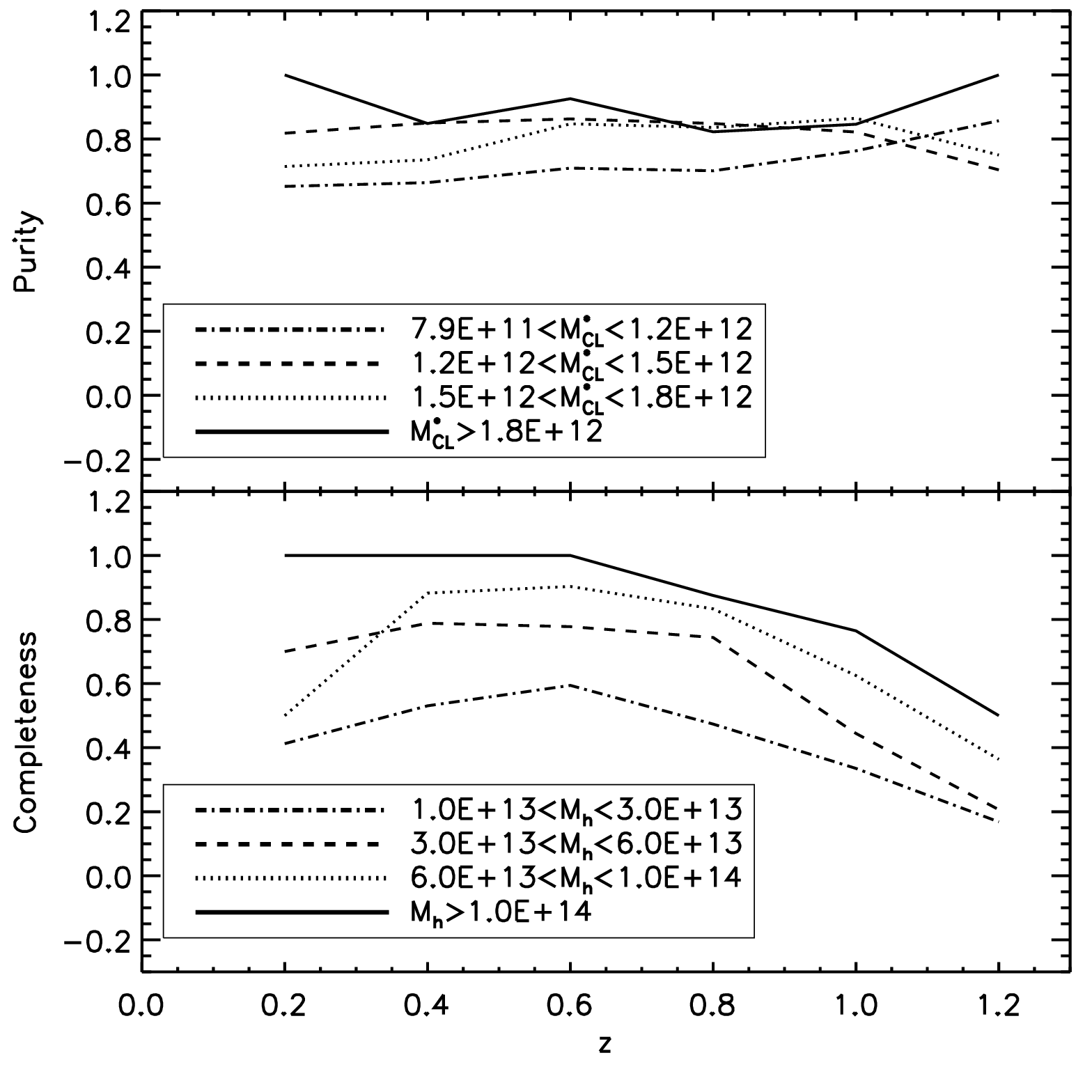} 
\caption{Purity (top panel) and completeness (bottom panel) rates as a function of redshift for different dark matter halo masses ($M_h$) and total stellar mass ($M^*_{\rm CL}$). The relationship between these two mass measurements is computed directly from the simulation and section \S4.3 is devoted to it. We find that the purity rates are always $>$60\% while the completeness rates decreases as a function of redshift, being $>$80\% (70\%) for total stellar masses $M^*_{\rm CL}>1.5\times10^{12}\rm M_{\odot}$ ($1.2\times10^{12}\rm M_{\odot}$) up to redshift 0.8. The selection function extracted from this analysis can be found in Table \ref{tab:tabSelF}.}
\label{fig:dratesmass}
\end{figure}

We have measured the rate of completeness, defined as the percentage of clusters detected out of the total simulated sample, and the purity, defined as the percentage of clusters simulated that were detected out of the total detected sample. In the bottom and top panel of Fig. \ref{fig:dratesmass}, we show these rates as a function of dark matter halo ($M_h$) and total stellar mass ($M^{*}_{\rm CL}$), respectively. The dark matter-total stellar mass relation has been directly calibrated from the simulation and we investigate it in the next subsection. 

While the purity rates remain $>$60\% for all redshift and mass richness, the completeness rates are lower and decrease with redshift in general. We obtain completeness and purity rates $>$80\% for clusters with total stellar masses larger than $1.5\times10^{12}\rm M_{\odot}$ or halo masses of  $6\times10^{13}\rm M_{\odot}$ up to redshift $\sim 0.8$. For higher redshifts ($0.8<z\le 1$), we increase the threshold mass for which we find completeness $>$80\% to $M^*_{\rm CL}>1.8\times10^{12}\rm M_{\odot}$, equivalently $M_{h}>1.0\times10^{14}\rm M_{\odot}$. We fail at detecting galaxy clusters with completeness and purity rates $>$80\%  for clusters at redshift $z>1$.

If we relax instead both completeness and purity rates to be higher than 70\%, we obtain a stellar mass limit of $1.2\times10^{12}\rm M_{\odot}$ or halo masses of  $3\times10^{13}\rm M_{\odot}$ up to redshift $\sim 0.85$, increasing to $1.5\times10^{12}\rm M_{\odot}$ or halo masses of  $6\times10^{13}\rm M_{\odot}$ between $0.85<z\le 0.95$ and to $1.8\times10^{12}\rm M_{\odot}$, equivalently $1.0\times10^{14}\rm M_{\odot}$, within $0.95<z\le 1.2$.

\begin{figure}
\centering
\includegraphics[clip,angle=0,width=1.\hsize]{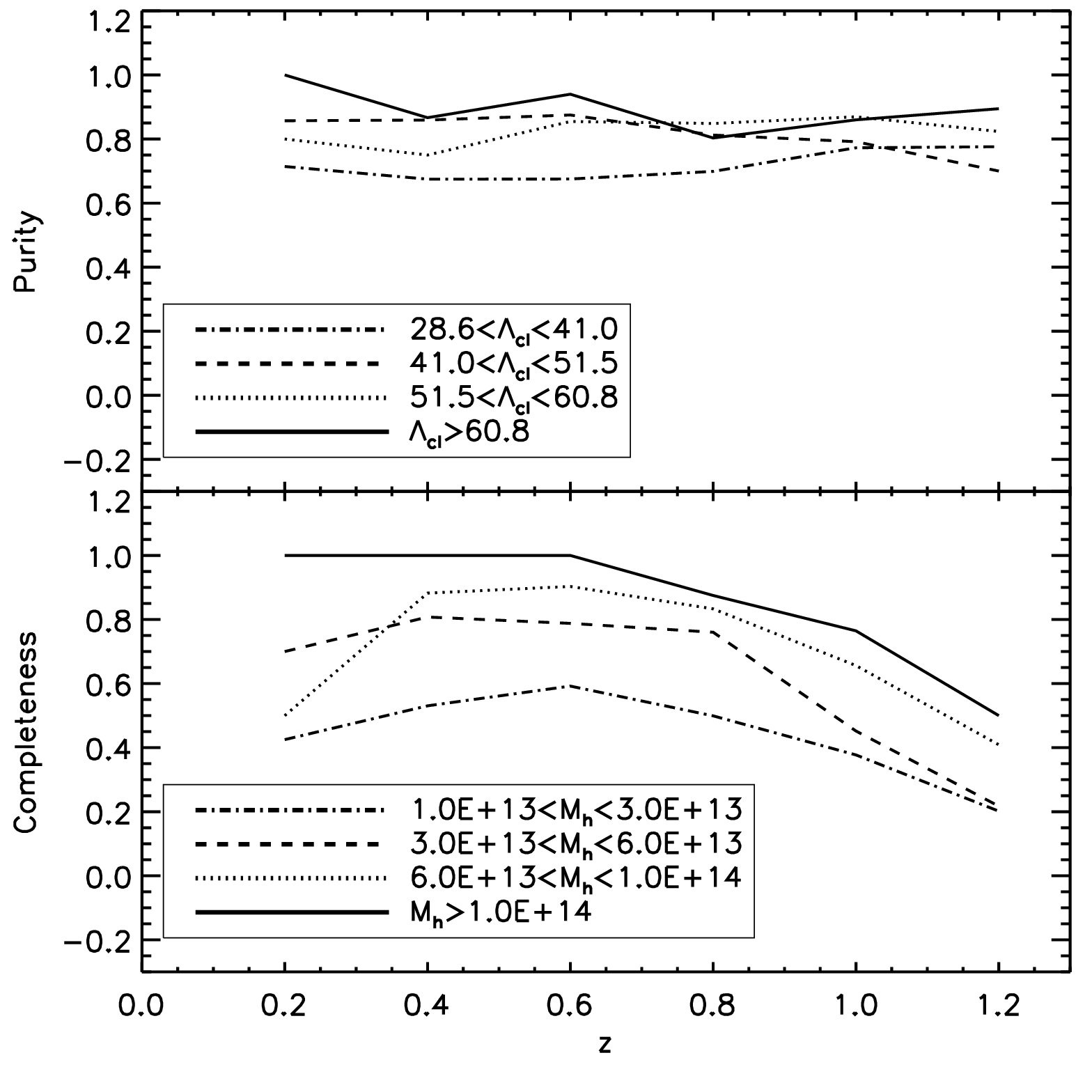} 
\caption{Purity (top panel) and completeness (bottom panel) rates as a function of redshift for different dark matter halo masses ($M_h$) and richnesses ($\Lambda_{\rm CL}$). The relationship between these two mass measurements is measured directly from the simulation and section \S4.3 is devoted to it. The purity rates are always $>$60\% while the completeness rates decreases as a function of redshift, being $>$80\% (70\%) for $\Lambda_{\rm CL}>51.5 (41)$ up to redshift 0.8. The selection function extracted from this analysis can be found in Table \ref{tab:tabSelF}}
\label{fig:dratelamda}
\end{figure}

We have examined the same rates as a function of redshift for different $\Lambda_{\rm CL}$ ranges, as shown in Fig. \ref{fig:dratelamda}. As before, the $\Lambda_{\rm CL}$ to dark matter calibration has also been measured from the mock catalogue. We see a very similar behaviour of the purity rates, being this $>$60\% for all redshift and mass ranges. Both completeness and purity become $>$80\% for $\Lambda_{\rm CL}>51.5$ up to redshift 0.8 and for $\Lambda_{\rm CL}>60.8$ for the redshift range $0.8 <z\le 1$. 

Decreasing the completeness and purity rates to 70\%, results into a lower $\Lambda_{\rm CL}$ threshold. We will be able to detect galaxy groups down to $\Lambda_{\rm CL}>41$ up to redshift 0.8, down to $\Lambda_{\rm CL}>51.5$ between redshift 0.85 and 0.95 and down to $\Lambda_{\rm CL}>60.8$ for redshifts higher than 1.

We have then summarized the selection function obtained from this analysis with three levels collected in Table \ref{tab:tabSelF}. The first level, includes the mass and redshift limits for our detections to be at least 80\% complete and pure and the second level refers to a level of, at least, 70\% of completeness and purity. We have also included a third level, of those detections that have a level of completeness higher than 50\% and purity higher than 60\%.

\begin{table}
      \caption{Selection functions for the ALHAMBRA Survey estimated from the full ALHAMBRA light-cone mock catalogue}
      \[
         \begin{array}{lcccc}
            \hline\noalign{\smallskip}
\multicolumn{1}{c}{\rm }&
\multicolumn{1}{c}{\rm z}&
\multicolumn{1}{c}{\rm M^*_{\rm CL}}&
\multicolumn{1}{c}{\rm \Lambda_{\rm CL}}&
\multicolumn{1}{c}{\rm M_{\rm h}}\\
\multicolumn{1}{c}{}&
\multicolumn{1}{c}{}&
\multicolumn{1}{c}{\rm (M_{\odot})}&
\multicolumn{1}{c}{}&
\multicolumn{1}{c}{\rm (M_{\odot})}\\
\hline\noalign{\smallskip}
 & 0.2-0.8  & >1.5\times10^{12} &>51.5 &>6\times10^{13} \\
\rm Level \, 1 \,(80\%) & 0.8-0.95  & >1.8\times10^{12} &>60.8 &>1\times10^{14} \\
& >0.95  & - & -& -\\
\hline
 & 0.2-0.85  & >1.2\times10^{12} &>41 &>3\times10^{13} \\
\rm Level \, 2 \,(70\%) & 0.85-0.95  & >1.5\times10^{12} &>51.5 &>6\times10^{13} \\
 & 0.95-1.2  & >1.8\times10^{12} &>60.8 &>1\times10^{14} \\
\hline
 & 0.2-0.8  & >7.9\times10^{11} &>28.6 &>1\times10^{13} \\
\rm Level \, 3 \, (50\%) & 0.8-0.95  &  -&- \\
& >0.95  & - & -& -\\
\hline
        \end{array}
      \]
\label{tab:tabSelF}
   \end{table}

\begin{figure}
\centering
\includegraphics[clip,angle=0,width=1.\hsize]{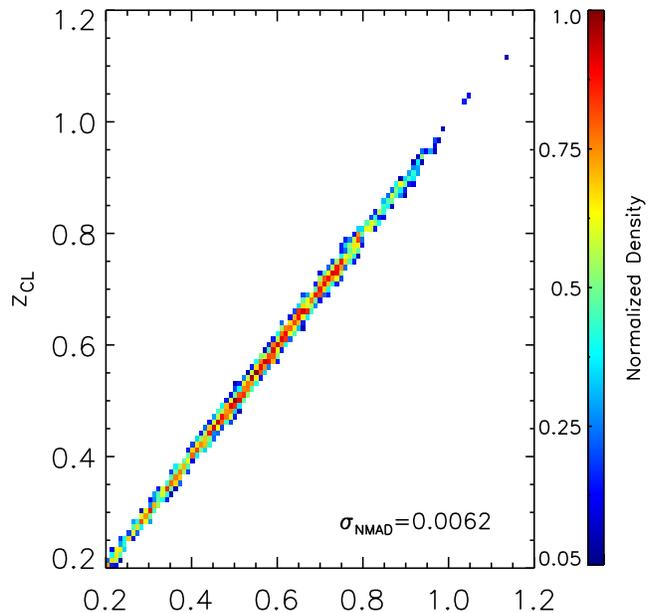} 
\caption{Colour-coded density map of the cluster recovered redshift versus input halo redshift for the matched detection in one of the ALHAMBRA mock catalogues. The dispersion obtained in the cluster redshift measurement is $\sigma_{\rm NMAD}\sim$0.0062, almost 2 times better than the photometric redshift precision of the photometric redshift of the survey.}
\label{fig:zinout}
\end{figure}

We have also investigated the accuracy with which we are able to recover the redshift of the clusters from the mock catalogue. In Fig. \ref{fig:zinout}, we show the halo input  redshift versus the recovered cluster redshift for those structures that are recovered.  We find an excellent agreement between both redshifts, achieving a dispersion, $\sigma_{\rm NMAD}=0.0062$, almost 2 times better than the mean photometric redshift accuracy of the ALHAMBRA survey (see \citealt{molino14} and section \S4.1).

\subsection{Dark matter halo mass - optical richness calibration}

The calibration of the mass  - observable relation for a cluster finder needs to be well understood in order to accomplish a realistic translation of the mass. This is particularly crucial for cosmological purposes with galaxy cluster counts (e.g. \citealt{rozo09}). We have measured the accuracy with which we can calibrate cluster and group masses with multiple medium-band photometry mimicking the ALHAMBRA data (section \S4.1). 

\begin{figure}
\centering
\includegraphics[clip,angle=0,width=1.\hsize]{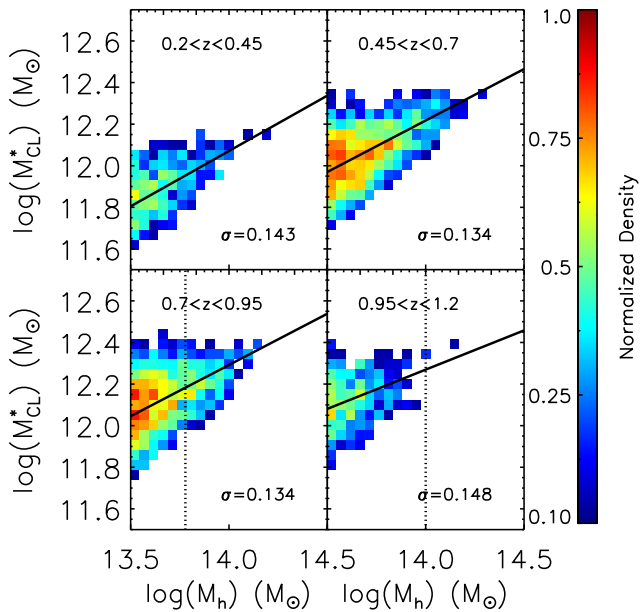} 
\caption{Density maps of the dark matter halo mass versus total stellar mass for the matched haloes in the ALHAMBRA mock catalogue for four different redshift bins. The solid line refers to the linear fit of the two quantities. The vertical dotted line shows the mass limit to which the fit is performed. The dispersion measured as the standard deviation between these two variables is displayed. for each different redshift bin.}
\label{fig:masssmass}
\end{figure}

In Fig. \ref{fig:masssmass}, we present the relation between the dark matter halo and total stellar mass richness for the matched output detections in the simulations for different redshift bins. These relations have been used to calibrate the observable total cluster stellar mass in Fig.\ref{fig:dratesmass}.

\begin{figure}
\centering
\includegraphics[clip,angle=0,width=1.\hsize]{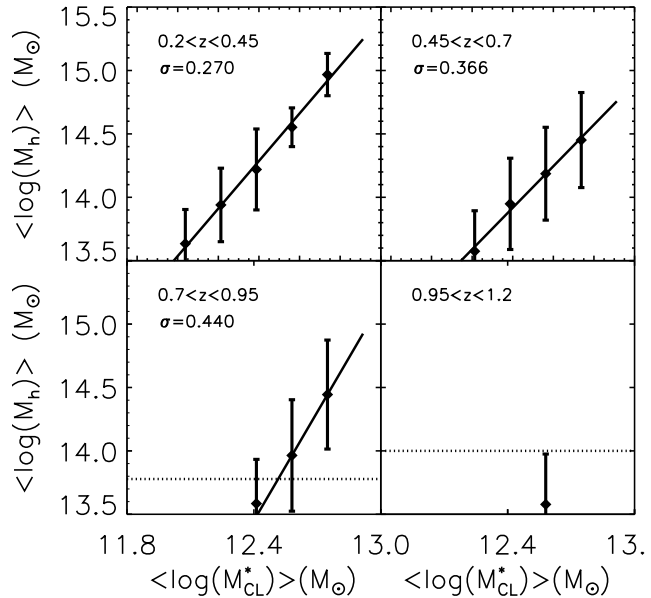} 
\caption{Average logarithm of the halo cluster mass as a function of the average logarithm of the total stellar mass for different redshift bins. The solid line refers to the linear fit of the two quantities. The dotted line shows the mass limit that we are complete and to which the fit is performed. The dispersion measured as the standard deviation between these two variables is displayed. for each different redshift bin. This dispersion becomes comparable to what has been found in other optical broad-band surveys down to 2-3 orders of magnitude higher mass limits and to other non-optical mass estimators such as X-ray luminosity or $Y_{SZ}$.}
\label{fig:smassmass}
\end{figure}

We now have calibrated the dispersion obtained given a particular richness (e.g $M^*_{CL}$ or $\Lambda_{\rm CL}$). To do this, we have used a Monte Carlo approach. For each richness value, we have sampled randomly 10000 times all the possible halo mass values available and obtained a mean value and scatter. The results are shown in Fig. \ref{fig:smassmass} for different redshift bins. We also quote the mean scatter, $\sigma_{M_h | M^*_{CL}}$, down to the mass limit we reaching for each redshift bin.

We find that the stellar mass has a very tight dispersion in recovering masses ($\sigma_{M_h | M^*_{CL}} \sim 0.27$ dex) down to the limit of $M_{\rm h}\sim 3\times10^{13}\rm M_{\odot}$. This value refers to the standard deviation between the two variables and it is similar to the values that present broad-band surveys have found for other optical proxies such as $N_{200}$, with the exception that we are able to sample  2-3 orders of magnitude lower limit in mass \citep{rozo09,hilbert10,andreon12}. In addition, this value is comparable to the precision that other non-optical proxies, such as X-ray luminosity or the total integrated SZ signal over the cluster, $Y_{SZ}$, are obtaining for a more extended mass range (e.g  \citealt{rozo14} and references herein). This result really demonstrates the enormous potential that multiple narrow and medium-band filters surveys have for calibrating cluster and group masses with excellent precision.

\begin{figure}
\centering
\includegraphics[clip,angle=0,width=1.\hsize]{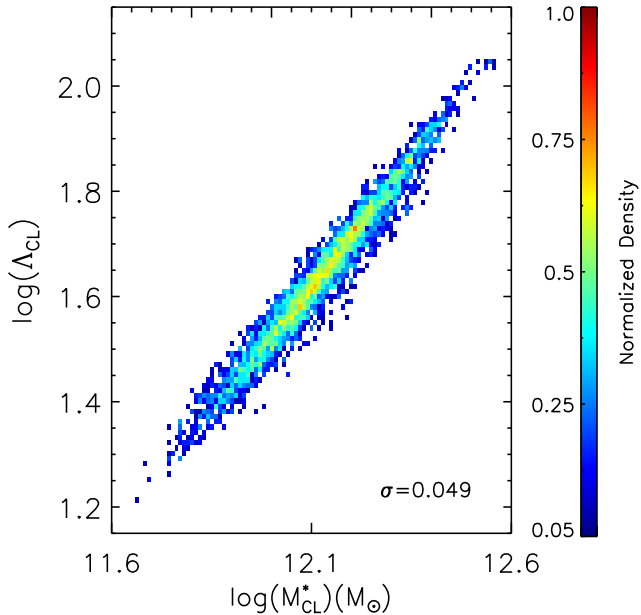} 
\caption{Density map of the logarithm of the  $M^*_{\rm CL}$ versus the $\Lambda_{\rm CL}$ parameter. We display the dispersion measured as the standard deviation between these two variables. As expected, there is a very tight relation linear between these two parameters, making their use almost irrelevant for calibrating the cluster halo mass.}
\label{fig:smasslamda}
\end{figure}

We have performed the same analysis with the  $\Lambda_{\rm CL}$ parameter, obtaining a very similar behaviour for the $M^*_{\rm CL}$, and a very similar dispersion. In order to show the proximity of these two parameters, we display in Fig.  \ref{fig:smasslamda}, the relation between the $\Lambda_{\rm CL}$ and the $M^*_{\rm CL}$ parameters. As expected, both parameters show a very tight linear relation with very low scatter (0.049). Hence, we choose to work with the $M^*_{\rm CL}$ hereafter.

\section{ALHAMBRA optical detections}

We have applied the BCF to the ALHAMBRA survey with the specifications mentioned before. As in the simulations, we have pre-selected the galaxies in the catalogue with $odds>0.5/(1+z)$. We have also removed from the catalogue those galaxies with saturated flags, $Satur\_Flag=1$ and high stellar indicator indices ($Stellar\_Flag>0.7$). Besides, we have used the masks determined in \cite{molino14}. 

We have set the three different thresholds determined in Table \ref{tab:tabSelF}. The number of cluster and groups corresponding to different levels of purity and completeness are 176, 359 and 365 for Level 1($>$80\%), Level 2($>$70\%) and Level 3 ($>$50\%) respectively. Note that even if Level 3 goes deeper in mass, it is more restrictive in redshift than Level 2 and 1, in order to achieve the 50\% completeness and purity rates. We have visually checked the Level 1 and 2 detections and eliminated eleven of these detections (3\%) that were centered on a flawed part of the image, obtaining a final sample of 171 and 348 structures for the Level 1 and 2, respectively. We have also included five additional structures detected at $z<0.2$ for inspection purposes. Then, for the Level 1 detections, we find  $\sim$ 66.2 clusters and groups per square degree in ALH-2 to 8 fields within $0.2 < z \le 0.95$, while for the Level 2 detections, the numbers increase to  $\sim$ 125.2 detections per square degree within $0.2 < z \le 1.2$. 

We include a list of the detections in Level 1 and 2 in Table \ref{tab:tabALH12} and the complementary detections to the Level 3 in Table \ref{tab:tabALH3}. The meaning of the columns are the following. The first column sets the name of the cluster or group detected,  the two following columns are the cluster center coordinates, the fourth column is the galaxy cluster redshift. The fifth and sixth columns refer to the total stellar mass, $M^*_{\rm CL}$, and the $\Lambda_{\rm CL}$ parameter respectively. In Table  \ref{tab:tabALH12}, we have set a last column indicating the level to which the detection belongs (Level 1 or 2). The catalogues will be available in the electronic version of the journal and can be also found online, together with a collection of colour images of the clusters\footnote{http://bascaso.net46.net/ALHAMBRA\_clusters.html}$^{,}$\footnote{http://alhambrasurvey.com/}. 

For the purpose of this work, we will use the detection in Level 2, as a compromise between high purity and completeness.

\begin{table*}
\centering
      \caption{Clusters and groups detected in the ALHAMBRA Survey with Level 1 and 2 of completeness and purity}
      \[
         \begin{array}{ccccccc}
            \hline\noalign{\smallskip}
\multicolumn{1}{c}{\rm Name}&
\multicolumn{1}{c}{\alpha (2000)}&
\multicolumn{1}{c}{\delta (2000)}&
\multicolumn{1}{c}{\rm z_{est}}&
\multicolumn{1}{c}{\rm M^*_{\rm CL}}&
\multicolumn{1}{c}{\rm \Lambda_{\rm CL}}&
\multicolumn{1}{c}{\rm Level}\\
\multicolumn{1}{c}{}&
\multicolumn{1}{c}{($deg$)}&
\multicolumn{1}{c}{($deg$)}&
\multicolumn{1}{c}{}&
\multicolumn{1}{c}{\rm (10^{12} M_{\odot})}&
\multicolumn{1}{c}{}&
\multicolumn{1}{c}{}\\
\hline\noalign{\smallskip}
\mbox{ALH0229.23+0108.13} & \mbox{ 02:29:23.30} & \mbox{+01:08:13.20} &     0.24 &         1.46 &        46.62 &     \rm    L2 \\
\mbox{ALH0229.55+0108.37} & \mbox{ 02:29:55.15} & \mbox{+01:08:37.32} &     0.25 &         1.64 &        52.46 &     \rm    L1 \\
\mbox{ALH0229.40+0108.47} & \mbox{ 02:29:40.15} & \mbox{+01:08:46.68} &     0.26 &         1.33 &        44.27 &     \rm    L2 \\
\mbox{ALH0227.13+0110.25} & \mbox{ 02:27:12.98} & \mbox{+01:10:25.32} &     0.38 &         1.35 &        41.73 &     \rm    L2 \\
\mbox{ALH0230.18+0105.18} & \mbox{ 02:30:18.00} & \mbox{+01:05:18.24} &     0.50 &         1.63 &        50.26 &     \rm    L1 \\
\mbox{ ALH0226.60+0107.2} & \mbox{ 02:26:59.98} & \mbox{+01:07:01.56} &     0.39 &         1.60 &        49.31 &      \rm   L1 \\
\mbox{ALH0226.56+0103.24} & \mbox{ 02:26:56.42} & \mbox{+01:03:23.76} &     0.60 &         1.55 &        48.60 &     \rm    L1 \\
\mbox{ ALH0226.46+0108.1} & \mbox{ 02:26:46.46} & \mbox{+01:08:00.60} &     0.38 &         1.25 &        39.17 &      \rm   L2 \\
\mbox{ ALH0228.16+0110.6} & \mbox{ 02:28:16.27} & \mbox{+01:10:06.24} &     1.17 &         2.00 &        70.75 &     \rm    L2 \\
\mbox{ALH0230.25+0103.50} & \mbox{ 02:30:25.03} & \mbox{+01:03:50.40} &     0.58 &         1.36 &        40.86 &    \rm     L2 \\
... & ...  & ...  & ...  & ...  & ...  & ... \\
\hline
         \end{array}
      \]
\begin{flushleft} 
Table \ref{tab:tabALH12} is available in the online version of the article. A portion is shown for illustration.
\end{flushleft}
\label{tab:tabALH12}
   \end{table*}

\begin{table*}
\centering
      \caption{Clusters and groups detected in the ALHAMBRA Survey with Level 3 of completeness and purity not included in Table \ref{tab:tabALH12}}
      \[
         \begin{array}{cccccc}
            \hline\noalign{\smallskip}
\multicolumn{1}{c}{\rm Name}&
\multicolumn{1}{c}{\alpha (2000)}&
\multicolumn{1}{c}{\delta (2000)}&
\multicolumn{1}{c}{\rm z_{est}}&
\multicolumn{1}{c}{\rm M^*_{\rm CL}}&
\multicolumn{1}{c}{\rm \Lambda_{\rm CL}}\\
\multicolumn{1}{c}{}&
\multicolumn{1}{c}{($deg$)}&
\multicolumn{1}{c}{($deg$)}&
\multicolumn{1}{c}{}&
\multicolumn{1}{c}{\rm (10^{12} M_{\odot})}&
\multicolumn{1}{c}{}\\
\hline\noalign{\smallskip}
\mbox{ ALH0229.54+0106.3} & \mbox{ 02:29:54.07} & \mbox{+01:06:03.24} &     0.26 &         1.02 &        33.30 \\
\mbox{ALH0229.10+0115.13} & \mbox{ 02:29:10.27} & \mbox{+01:15:12.60} &     0.23 &         0.81 &        24.74 \\
\mbox{ ALH0227.9+0103.46} & \mbox{ 02:27:09.31} & \mbox{+01:03:46.08} &     0.36 &         1.16 &        36.38 \\
\mbox{  ALH0229.5+0108.1} & \mbox{ 02:29:05.38} & \mbox{+01:08:01.32} &     0.38 &         1.08 &        34.87 \\
\mbox{ALH0228.16+0114.46} & \mbox{ 02:28:15.62} & \mbox{+01:14:46.32} &     0.39 &         0.79 &        24.38 \\
\mbox{ALH0227.56+0104.22} & \mbox{ 02:27:55.66} & \mbox{+01:04:22.08} &     0.39 &         0.90 &        28.11 \\
\mbox{ALH0230.14+0112.47} & \mbox{ 02:30:14.04} & \mbox{+01:12:46.80} &     0.69 &         0.92 &        31.57 \\
\mbox{ALH0226.47+0112.49} & \mbox{ 02:26:47.42} & \mbox{+01:12:49.32} &     0.64 &         0.89 &        28.39 \\
\mbox{ALH0228.56+0039.10} & \mbox{ 02:28:55.75} & \mbox{+00:39:10.44} &     0.26 &         0.91 &        27.77 \\
\mbox{ALH0228.32+0044.37} & \mbox{ 02:28:31.82} & \mbox{+00:44:37.32} &     0.22 &         0.98 &        30.18 \\
... & ...  & ...  & ...  & ...  & ...  \\
\hline
         \end{array}
      \]
\begin{flushleft} 
Table \ref{tab:tabALH3} is available in the online version of the article. A portion is shown for illustration.
\end{flushleft}
\label{tab:tabALH3}
   \end{table*}
   
\subsection{Comparison with other studies}

One of the most attractive features of the ALHAMBRA survey is the number of different surveys with overlapping data, providing an excellent way to deal with the cosmic variance (e.g. \citealt{molino14,lopez-sanjuan14}). In addition, we can take advantage of early searches on these fields to cross-correlate them with our detections and quantify the degree of agreement between different datasets and methodologies.

From all the surveys extending over the seven ALHAMBRA fields, one of them has been widely studied for cluster purposes in the literature: the COSMOS survey, overlapping with the ALH-4 field. Up to date, several work has provided us with catalogues of large-scale structure or cluster/group detections in the COSMOS survey in the optical.  \cite{scoville07b} detected very large-scale structure up to redshift 1.1 in the COSMOS field using photometric redshifts. In addition, \cite{olsen07} detected clusters in the CFHTLS-Deep Survey, also overlapping with the COSMOS field. Later on, \cite{bellagamba11} used an optical and weak lensing search to detect galaxy clusters in the COSMOS field. Two recent works,  \cite{castignani14} and \cite{chiang14} have provided a list of high redshift ($z>1$) clusters and proto-clusters detections. The former is based on detecting over densities around radio galaxies to detect high-redshift ($z\sim1-2$) galaxy clusters and the latter, used the photometric redshift obtained from the combination of the optical and IR data to detect very high-redshift proto-clusters  ($z\sim2-7$).

In addition, there are some more works that have been detecting groups and clusters with other techniques: 
\cite{knobel09} used a combination of a Friends-of-Friends (FoF) and Voronoi-Delaunay Method (VDM) algorithms to detect clusters optically using the zCOSMOS spectroscopic  sample and   \cite{finoguenov07} have performed an extended search of galaxies in clusters in the 36 XMM-Newton pointings on the COSMOS field,  obtaining calibrated weak lensing masses  \citep{leauthaud10}. 

\begin{figure*}
\centering
\includegraphics[clip,angle=0,width=1.0\hsize]{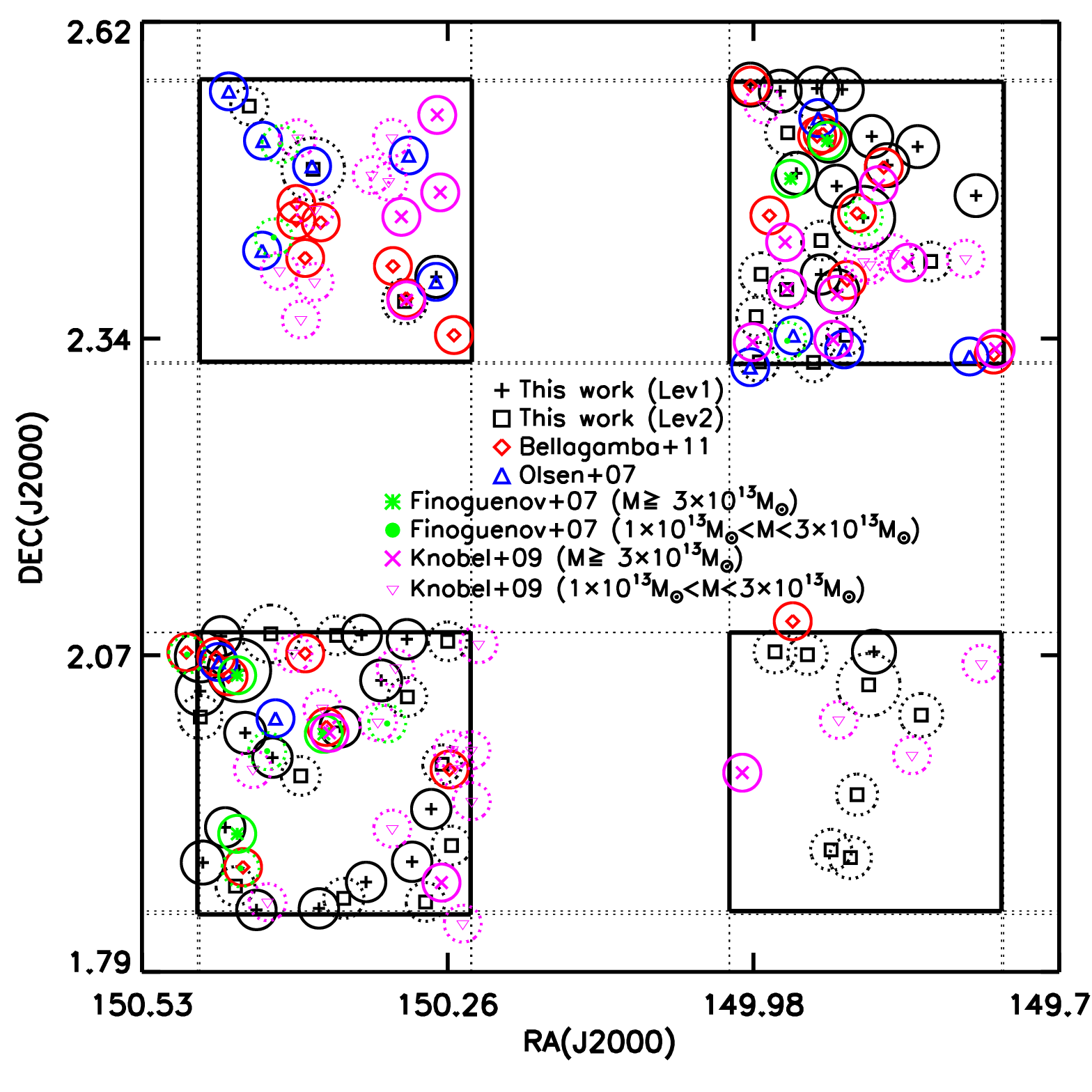} 
\caption{Spatial distribution of the cluster and group detections in the ALH-4 field. The black plus symbols (and solid circles) and square symbols (and dotted circles) refer to the detections found in this work following the Level 1 and 2 selection respectively. The red diamonds and blue triangles  refer to the optical detections found by Bellagamba et al. 2011 and Olsen et al. 2007 respectively. The green asterisk (and solid circle) and green crosses (and dotted circle) allude to the X-ray detections found by Finoguenov et al. 2007 with estimated masses $M\ge 3 \times 10^{13} \rm M_{\odot}$ and $1 \times 10^{13} \rm M_{\odot}<M<3 \times 10^{13} \rm M_{\odot}$, respectively. Finally, the magenta asterisks (and solid circle) and pink crosses (and dotted circle) make reference to the spectroscopic detections found by Knobel et al. 2009 with estimated masses $M\ge 3 \times 10^{13} \rm M_{\odot}$ and $1 \times 10^{13} \rm M_{\odot}<M<3 \times 10^{13} \rm M_{\odot}$, respectively. The size of the circle corresponds to a 500 Kpc radius sphere at the redshift of the cluster. The solid lines define the four discontiguous fields of ALH-4 (see Fig. A.1 in Molino et al. 2014 for a description of the geometry of the survey). }
\label{fig:ALH4all}
\end{figure*}

We have compared our detections with the optical detections by \citealt{bellagamba11}, (B11) and  \citealt{olsen07} (O07) catalogues since they search for individual detections in the same redshift range as we do. We have also performed a comparison with the X-ray detections found by  \citealt{finoguenov07} (F07) and with the spectroscopic detections found by \citealt{knobel09} (K09) in zCOSMOS, only considering those clusters with $M>3 \times 10^{13} \rm M_{\odot}$. In Fig. \ref{fig:ALH4all}, we show the spatial distribution of all these detections in the ALH-4 field, consisting of 4 separated regions. For illustration purposes, we have included the detections found by F07 and K09 within $1\times 10^{13} \rm M_{\odot} \le M \le 3 \times 10^{13} \rm M_{\odot}$.

We do not expect an exact distribution of the detections, since their selection functions are built from different surveys with different depths and sets of data and using different methods, with different systematics. However, we find a good agreement  between some of the structures found in the field by eye. For instance, it is well-known that the COSMOS field has two main large-scale structures (e.g. \citealt{scoville07a,molino14}), which are basically recovered by all the works. We also find a basically `empty' subfield, in agreement with all the other methods. We analyse in more detail the general level of agreement below.

We have first matched the ALHAMBRA detections to the four different studies in order to study the level of agreement of the different samples. We have used an analogous FoF algorithm described in \cite{ascaso12}, which we summarise here briefly. We first make a list of a friends of friends of every candidate, where `friends' are defined as all the detections found within 1 Mpc. Then, we restrict this candidate list to those `friends' whose photometric redshift satisfy Eq. \ref{eq:zdiff}. Finally, we select the closest `friend' as the best match.

We have obtained 86.36\%, 84.61\%, 100\% and 73.30\% of agreement with the B11, O07, F07 and K09 samples respectively. Note that these values confirm the completeness rates found in the simulations. In addition, the accuracy with which we recover the main redshift of the cluster with respect to the redshift found by the four works is $0.021 \pm 0.032$ (B11), $0.130 \pm 0.167$ (O07), $-0.007 \pm 0.0095$ (F07) and $0.018 \pm 0.018$ (K09).

\begin{figure}
\centering
\includegraphics[clip,angle=0,width=1.0\hsize]{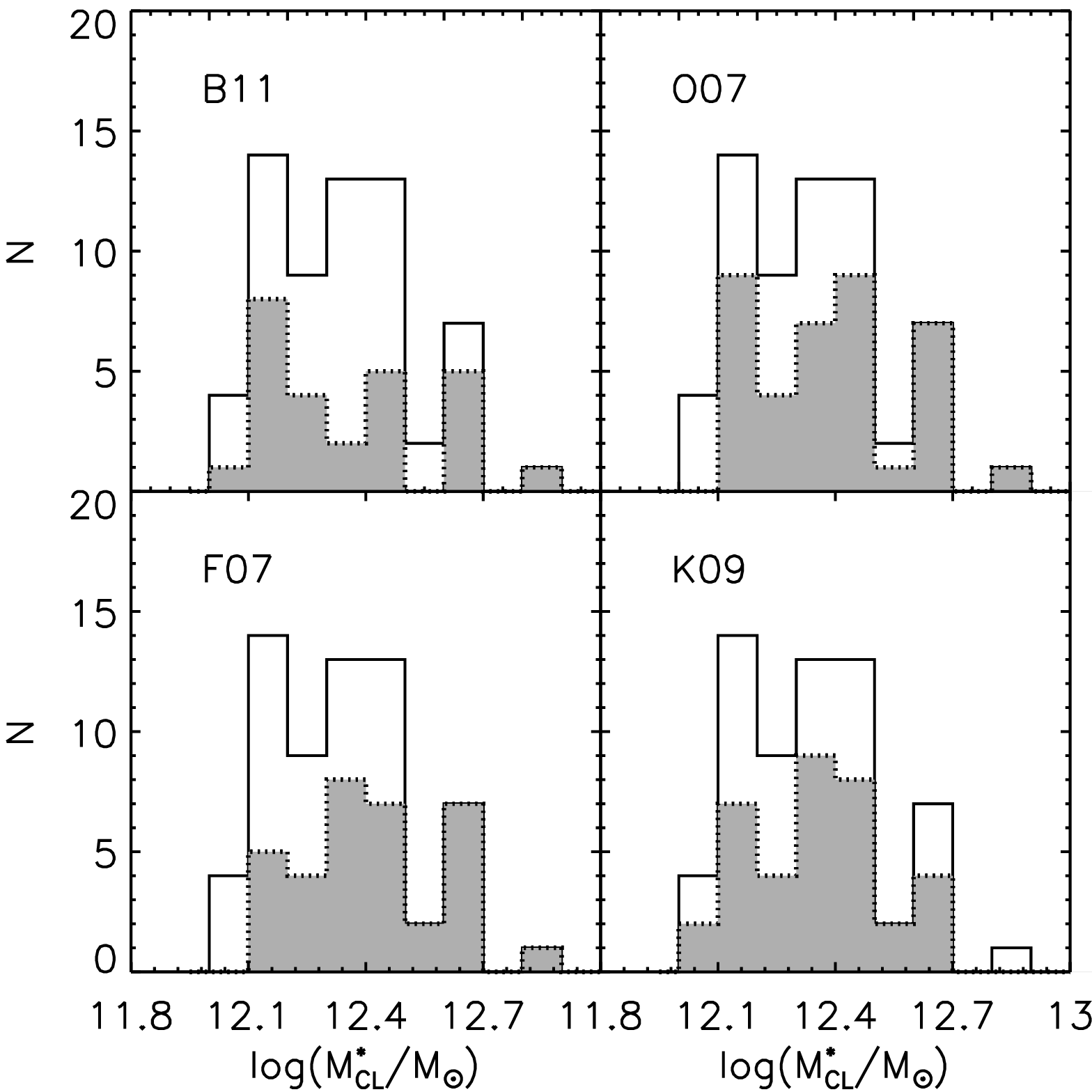} 
\caption{Total stellar mass distribution for the ALH-4 field (solid line) overplotted with the total stellar mass distribution of the ALH-4 subsample that has a counterpart in each of the considered works (dotted line and shaded area; Bellagamba et al. 2011 optical/weak-lensing, Olsen et al. 2007 CFHTLS, Finoguenov et al. 2007 X-ray, and Knobel et al. 2009 zCOSMOS samples respectively). The more massive structures are well matched for the O07, F07 and K09 samples, finding  a departure in the detections at lower masses. For B11, the rate of agreement is constant within the mass range.}
\label{fig:SMcomp}
\end{figure}

We have also examined the agreement between  the detections found in other work with respect to the ones found in this study. In Fig. \ref{fig:SMcomp} we show the distribution of total stellar mass of the clusters detection in ALH-4. The shaded area histogram shows the distribution of the total stellar mass of those detections that have a counterpart in each of the different catalogues. We notice a very good agreement of the more massive structures ($M_{\rm h}>1\times 10^{14}M_{\odot}$, equivalent to $M^*_{\rm CL}>1.8\times 10^{12}M_{\odot}$) for K09 ($>$64\%) and for O07 and F07 ($>$70\%) while a departure of the distribution at lower masses. As for the comparison with B11, we do not find a particular better agreement at high or low-masses being within the whole mass range  $\sim$50\%.

\begin{figure}
\centering
\includegraphics[clip,angle=0,width=1.0\hsize]{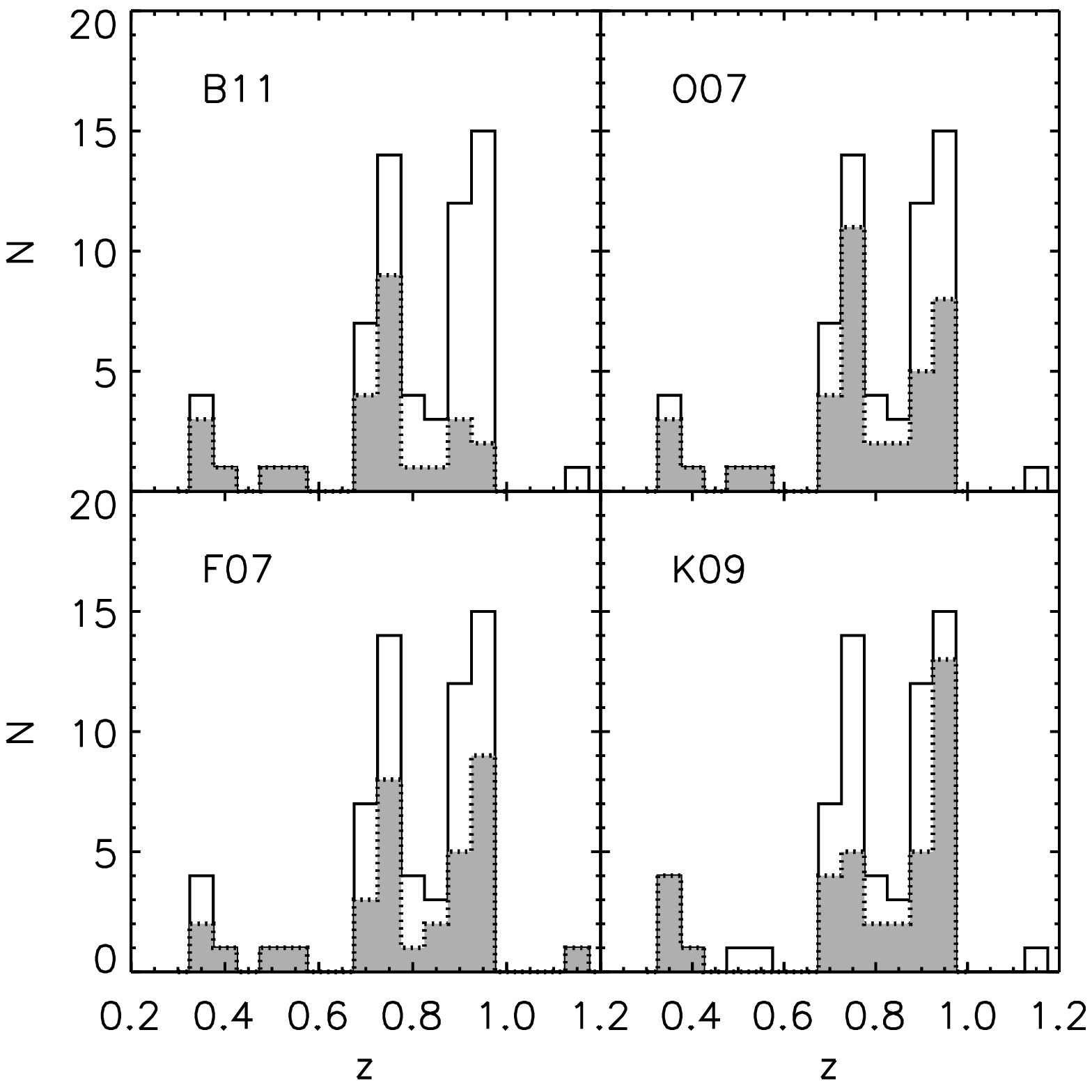} 
\caption{Redshift distribution for the ALH-4 field (solid line) overplotted with the redshift distribution of the ALH-4 subsample that has a counterpart in each of the different works considered (dotted line and shaded area; Bellagamba et al. 2011, Olsen et al. 2007, Finoguenov et al. 2007 and Knobel et al. 2009 respectively). We find an overall  good trace of the redshift range up to redshift 0.7 for B11, O07 and F07 finding departures at higher redshift range. The behavior is different for the K09 sample, finding a good agreement at z$<$0.5 and z$>$0.85. More details are given the text.}
\label{fig:zcomp}
\end{figure}

Similarly, in Fig. \ref{fig:zcomp}, we show the distribution of redshift of the detections in ALH-4, where the dotted line and shaded area shows the distribution of the redshift only for those detections that have a counterpart in each of the different catalogues. For the optical works, B11 and O07, we find a good agreement ($>$70\%) up to redshift 0.7 and we find a departure of the distribution at higher redshifts. For the X-ray group catalogue by F07, we find an agreement of $\sim$ 60\% up to redshift 0.7, decreasing slightly at higher redshift. Finally, for the spectroscopic group sample by K09, we do find a very good agreement up to redshift 0.5 ($>$80\%) and at redshift higher than 0.85 ($>$65\%), whereas the agreement is worse within $0.5\le z \le 0.85$ ($>$43\%).

\begin{figure}
\centering
\includegraphics[clip,angle=0,width=1.0\hsize]{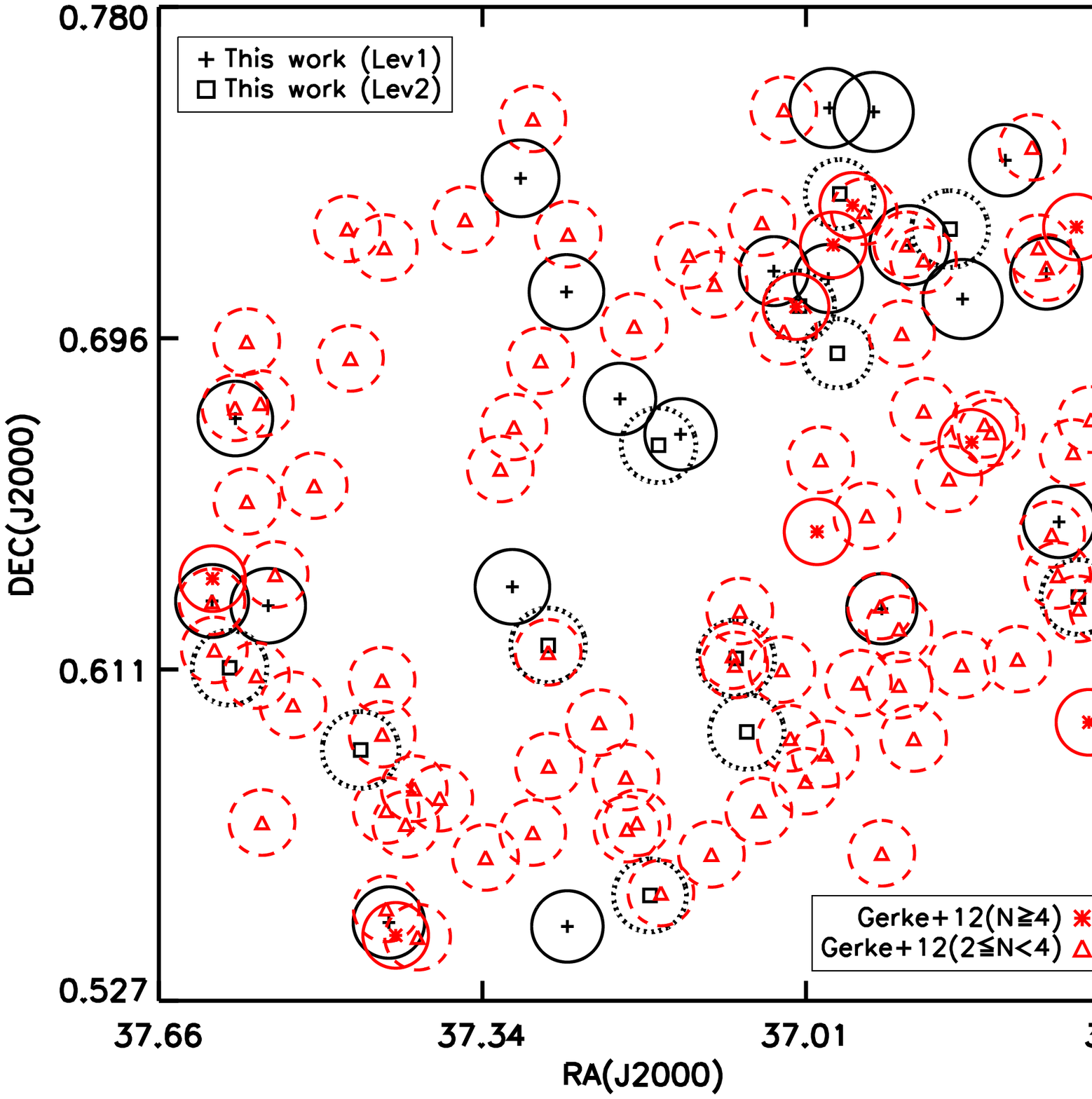} 
\caption{Spatial distribution of the cluster and group detections in one of the stripes of the ALH-2 field (0.25 $deg^2$). The black plus symbols (and solid circle) and square symbols (and dotted circle) refer to the detections found in this work restricted to the redshift range $0.65<z<1$ for the Level 1 and 2 selection respectively.  The red asterisks (and solid circle) and the red triangles (and dashed circle) refer to the detections found by Gerke et al. 2012 at redshift $<$1 with a minimum number of members of 4, and between 2 and 4 members, respectively.  The size of the circle corresponds to a 1 Mpc radius sphere at the redshift of the cluster.}
\label{fig:ALH2all}
\end{figure}

In addition to ALH-4, ALH-2 also overlaps with DEEP2 and \cite{gerke12} have performed a group search using the VDM. Their catalogue contains several groups with two or more members over the redshift range $0.65<z<1.5$. We have compared our detections with the \citealt{gerke12}, (G12) group catalogue, restricted to detections with at least three members. In Fig. \ref{fig:ALH2all}, we show the spatial distribution of all these detections in the stripe of the ALH-2 field (0.25 deg$^2$) which overlaps with DEEP2. Both samples have been restricted to the redshift range $0.65<z<1$ in order to compare complete samples in a similar range of redshift. 

We see an overall agreement between some of the structures found in G12 and in our sample. Indeed, we find that 60\% of the structures that G12 finds are recovered with a main redshift difference of $z_A-z_G=-0.001 \pm   0.021$, being $z_A$ and $z_G$, the redshift of this work and G12, respectively. This agreement increases to 75\% if we consider those structures with at least five members, obtaining a redshift difference of $z_A-z_G=0.001 \pm   0.0015$. 

\begin{figure}
\centering
\includegraphics[clip,angle=0,width=1.0\hsize]{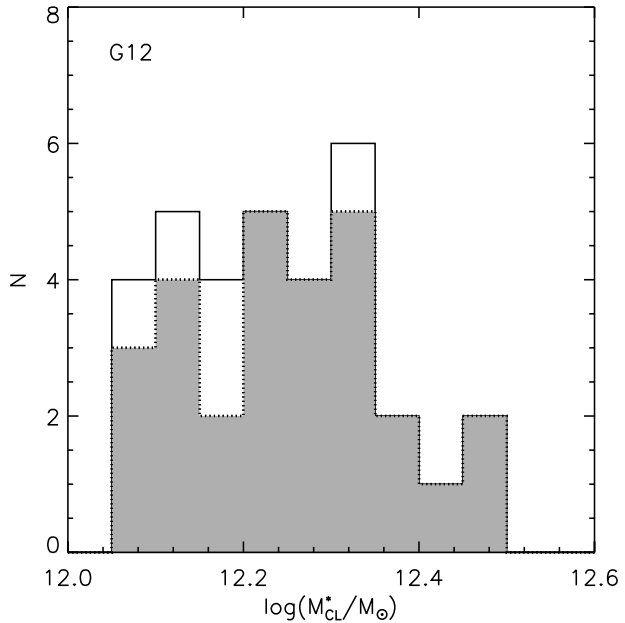} 
\caption{Total stellar mass distribution for the stripe of the ALH-2 field that overlaps with DEEP2 (solid line) overplotted with the total stellar mass distribution of the ALH-2 subsample that has a counterpart in the sample of Gerke et al. 2012. A very good agreement is noticeable (84.84\%), particularly in the more massive end of the distribution.}
\label{fig:SMcomp2}
\end{figure}

We have also noticed some other structures detected in this work that are not recovered by G12. In order to investigate which kind of detections G12 are recovering from our sample, we show in Fig. \ref{fig:SMcomp2}, the total stellar mass distribution of our restricted detections and overplotted, the stellar mass distribution only for those structures with a counterpart in the G12 sample. We see that both distributions agree very well finding a general agreement of 85\%, being slightly higher ($>$93\%) for high-mass clusters ($>7.0\times10^{13}M_{\odot}$). We then confirm that the BCF applied to ALHAMBRA-like surveys is able to recover low-mass groups in a similar way as spectroscopic surveys.

Complementary, ALH-3 and -8 also overlap with the Sloan Digital Sky Survey (SDSS). Nevertheless, the limiting area of ALHAMBRA restricts its detection to redshift $>$0.2, whereas the bulk of the SDSS detections are below this redshift.

These comparisons suggest that the optical methods are able to recover the redshift of the cluster similarly to the BCF, and with a similar precision as the spectroscopic sample, at least at lower ($z<0.5$) and higher redshift ($z>0.65$). On the other hand, the X-rays and spectroscopic methods seem to estimate  the cluster mass similarly to the BCF (70\%), whereas two out of the three optical methods trace similarly the mass distribution obtained with the BCF. 

The different cluster and group catalogues analysed in this work are built from different datasets, using a variety of methodologies each of them carrying their own systematics,  and with different selection functions. These facts make the comparison difficult, as it becomes evident in Fig. \ref{fig:ALH4all}, where the different spatial distributions of each dataset become noticeable. Notwithstanding these facts, we still find a general tendency in this comparison to agree at the high-end of the mass distribution function ($M_{\rm h}>1\times 10^{14}M_{\odot}$ at least) and at redshifts ($z<0.7$ at least), as expected from the simulations.

\subsection{Properties of the ALHAMBRA groups}

As mentioned in \S1, low-mass group samples within a relatively wide area of the sky are scarce and tend to suffer from observational biases. In the next subsections we review basic properties of the detections, we provide examples of such detections and we compare these results with previous studies. 

In this work, we do not intend to perform an exhaustive analysis on the properties of the galactic population of these structures. Rather, we mean to illustrate the variation of the properties of the low-mass groups with respect to the high-mass clusters.

\subsubsection{Stellar mass distribution}

\begin{figure*}
\centering
\includegraphics[clip,angle=0,width=0.45\hsize]{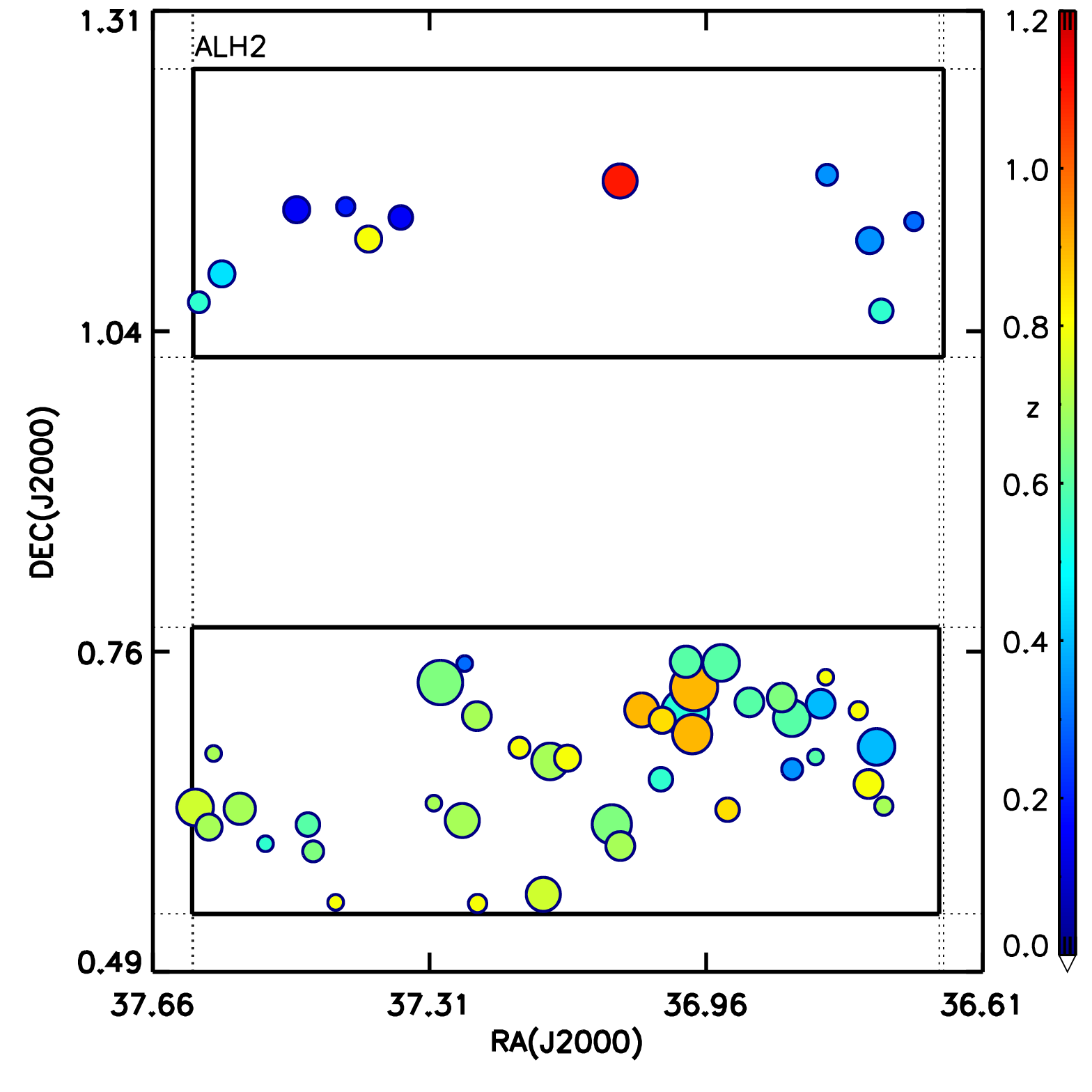} 
\includegraphics[clip,angle=0,width=0.45\hsize]{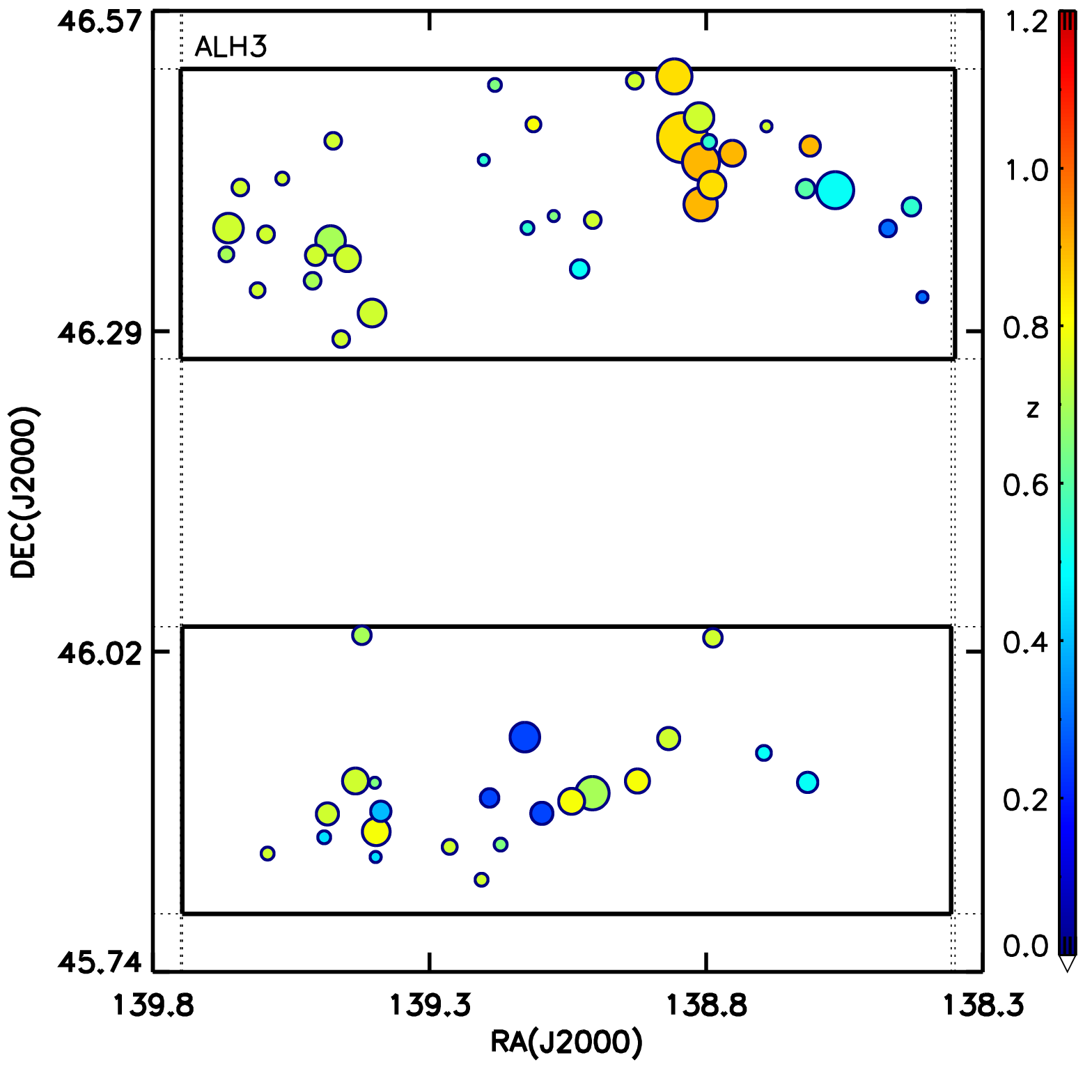} 
\includegraphics[clip,angle=0,width=0.45\hsize]{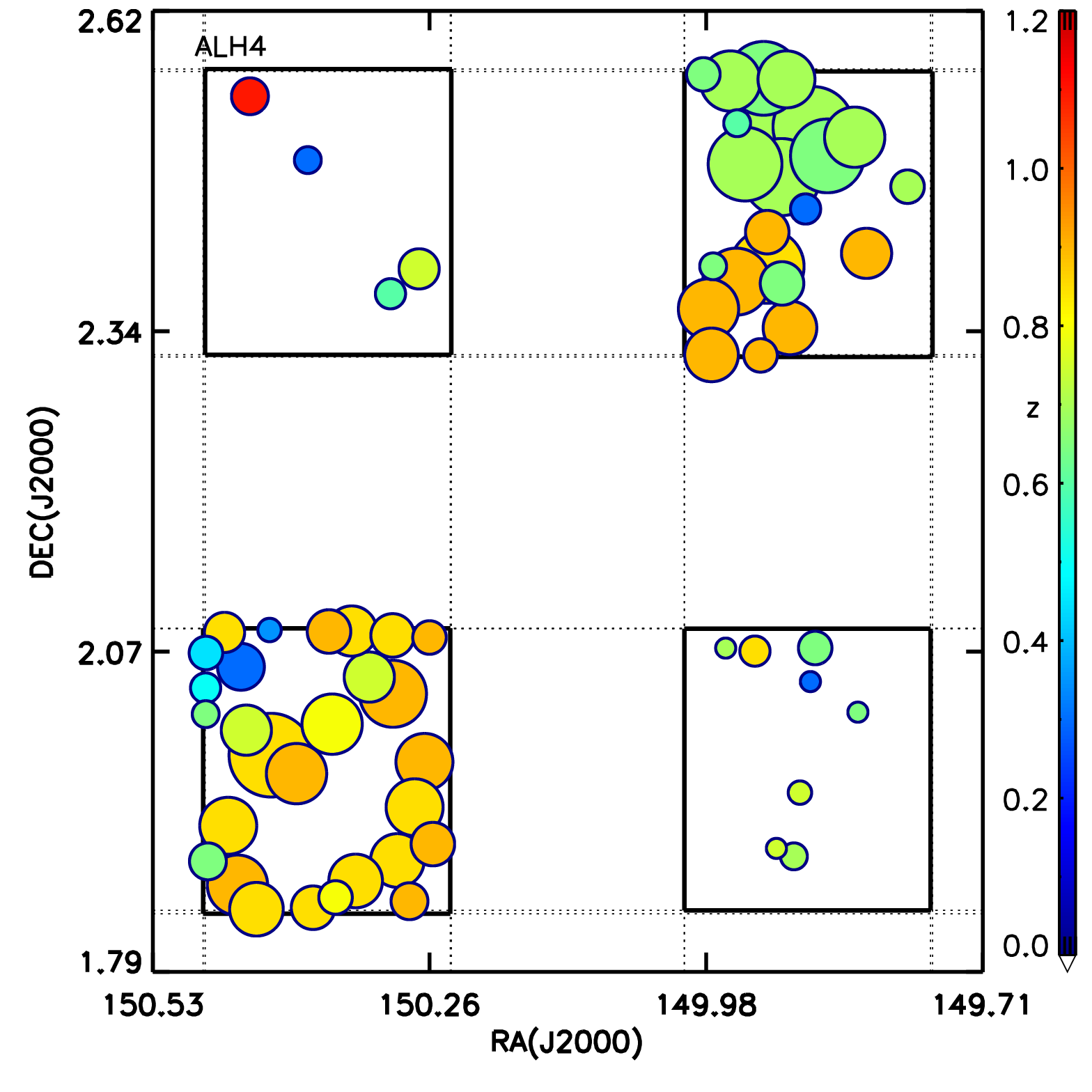} 
\includegraphics[clip,angle=0,width=0.45\hsize]{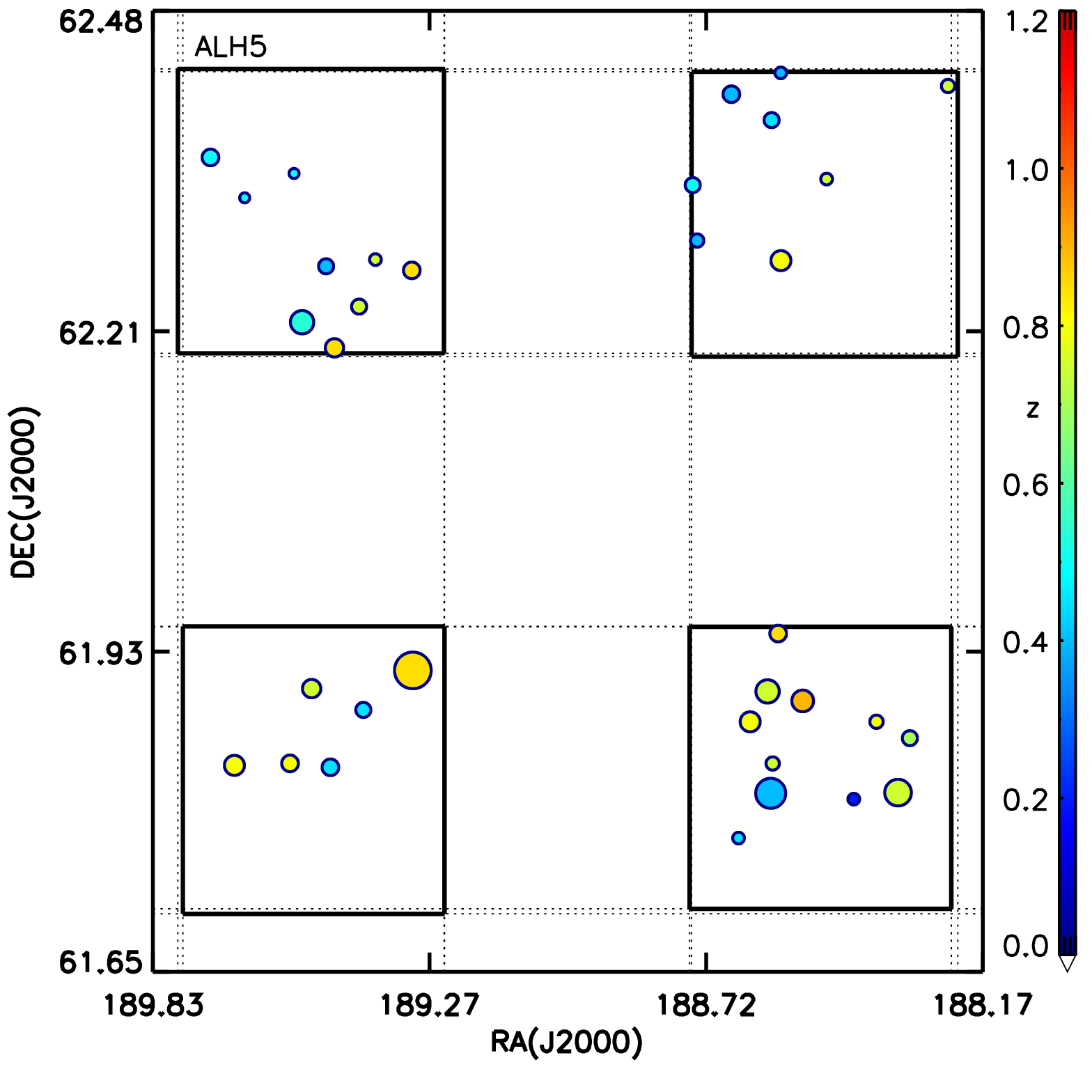} 
\caption{Spatial distribution of the detections in ALH-2, ALH-3, ALH-4 and ALH-5 fields (top left, top right, bottom left and bottom right panel respectively) for the Level 2 detections. The size of the circle is directly proportional to the measured total stellar mass and the colour scale refers to the redshift at which the redshift is located. The solid lines define the limits of each of the fields (see Fig. A.1 in Molino et al. 2014 for a description of the geometry of the survey). The same scale applies to all the panels.}
\label{fig:spatialdist}
\end{figure*}

\begin{figure*}
\centering
\includegraphics[clip,angle=0,width=0.45\hsize]{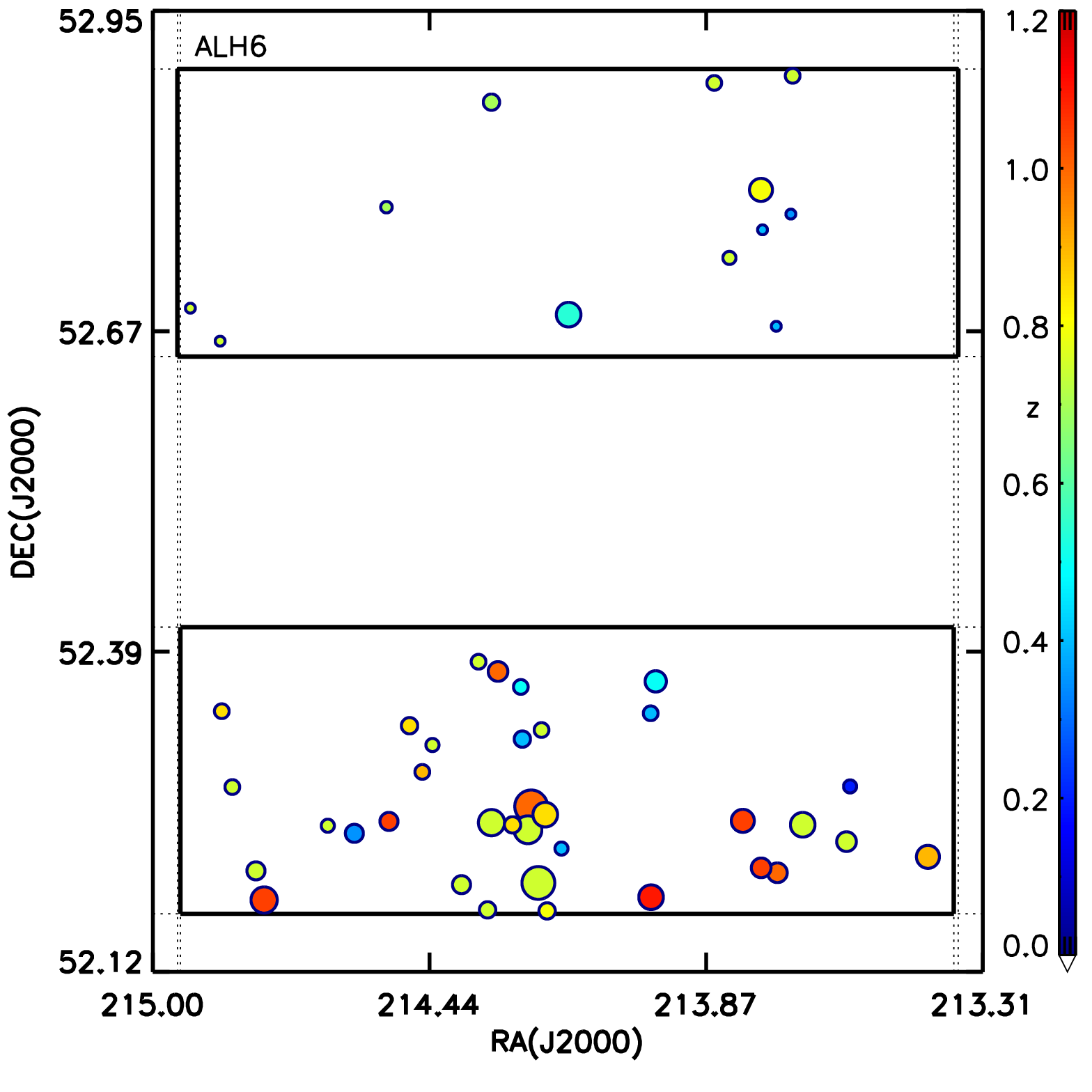} 
\includegraphics[clip,angle=0,width=0.45\hsize]{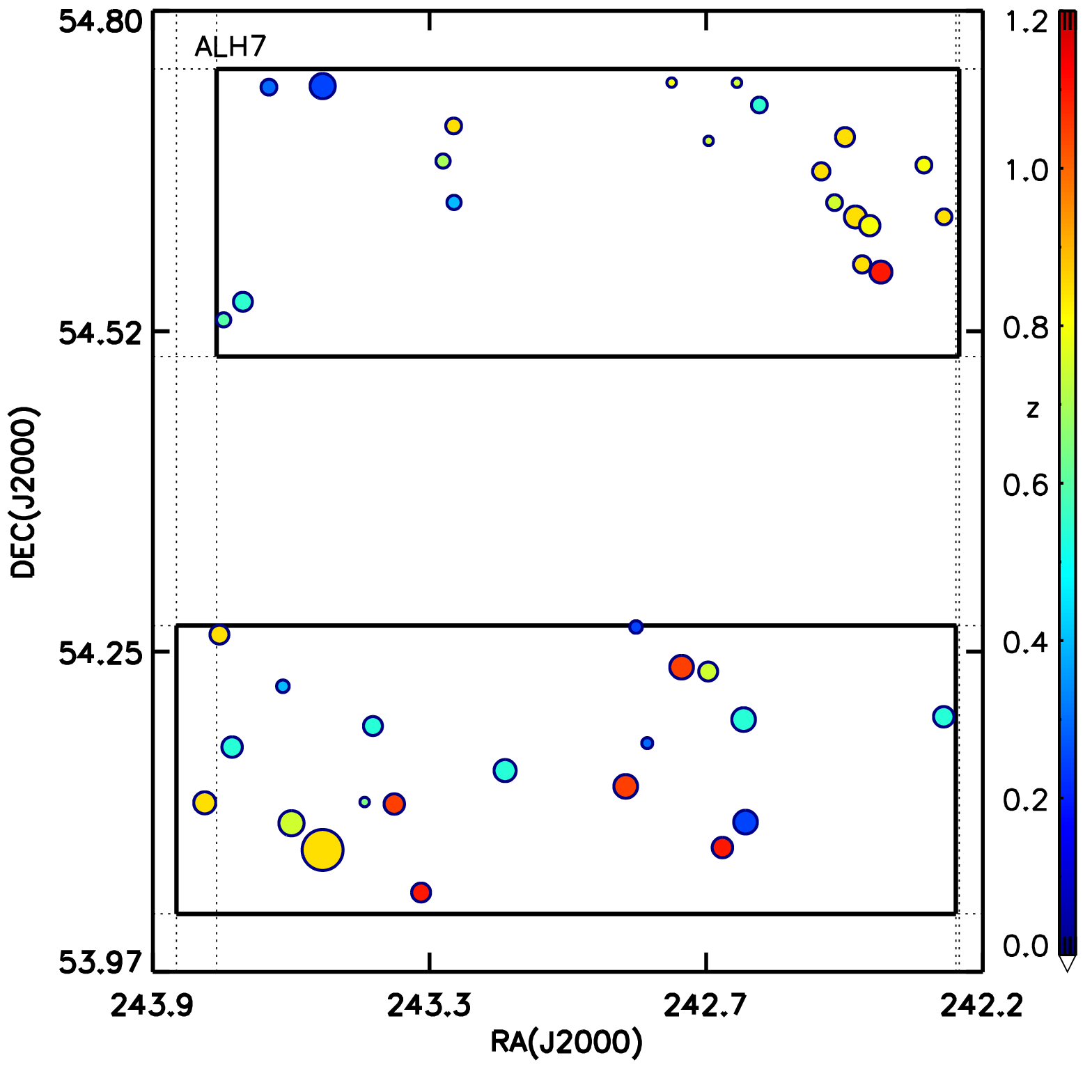} 
\includegraphics[clip,angle=0,width=0.45\hsize]{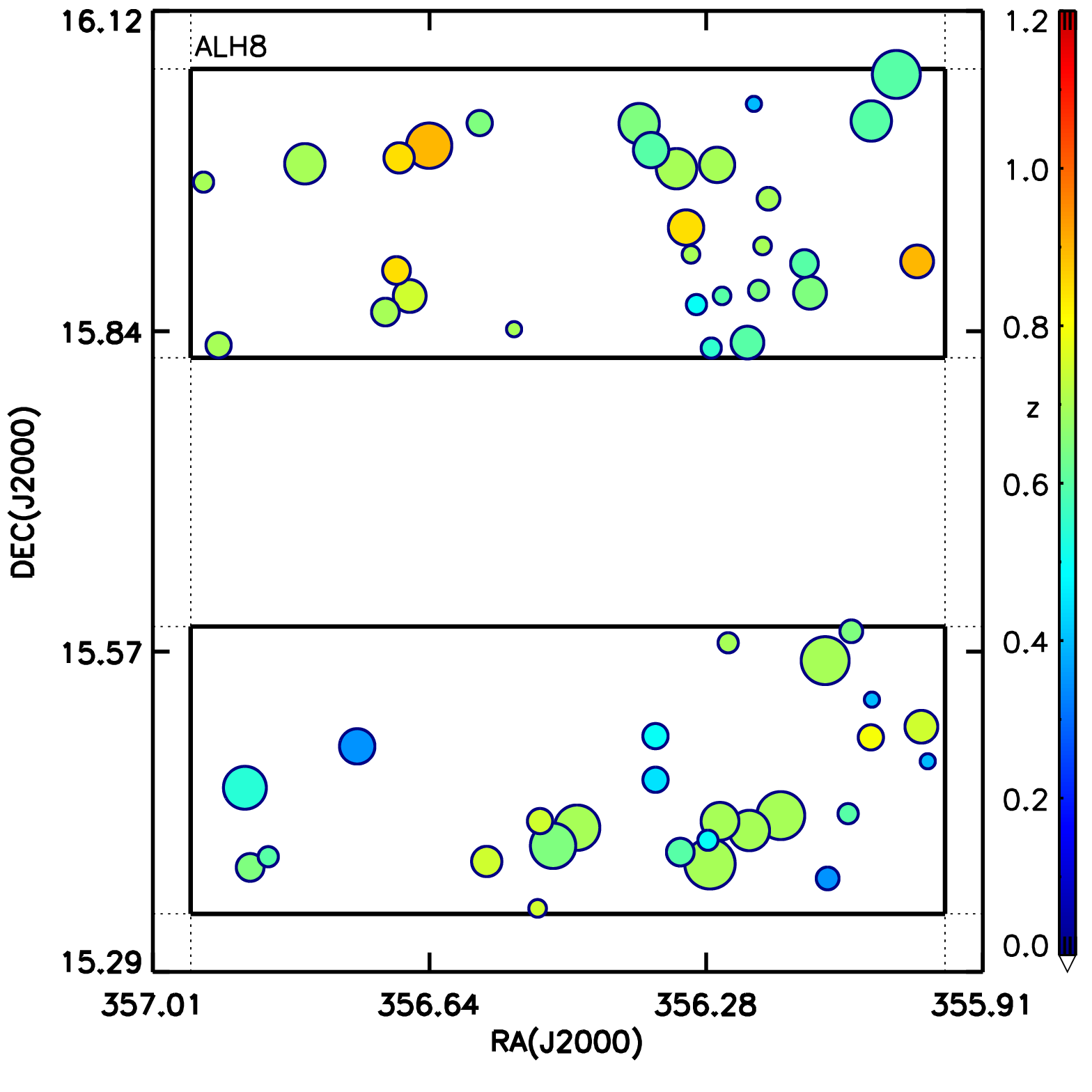} 
\caption{Spatial distribution of the detections in ALH-6, ALH-7 and ALH-8 fields (top left, top right and bottom central panel respectively) for the Level 2 detections. The size of the circle is directly proportional to the measured total stellar mass and the colour scale refers to the redshift at which the redshift is located. The solid lines define the limits of each of the fields (see Fig. A.1 in Molino et al. 2014 for a description of the geometry of the survey). The same scale applies to all the panels.  The different distributions between the fields is noticeable. The ALH-4 field becomes the densest and one of the most massive, while other fields as ALH-5 or ALH-7 are populated with less massive clusters.}
\label{fig:spatialdist2}
\end{figure*}

In Figs. \ref{fig:spatialdist} and  \ref{fig:spatialdist2}, we show the spatial distribution of the detections found in each of the fields of ALHAMBRA, proportionally scaled to the total stellar mass and colour-coded by their redshift. This distribution provides a visual way to study the large scale structure and filaments together with their cosmic variance. It becomes noticeable that the ALHAMBRA fields have very different large-scale structures. Fields, like ALH-6 or ALH-7 are significantly empty, compared with others such as ALH-4 which displays very massive structures. In fact, the well-known filaments found in COSMOS at $\sim$ 0.4, 0.7 and 0.9 \citep{scoville07b} are patent in this figure.  

It is also interesting to see that some fields, like ALH-5 and ALH-7 have significantly less massive clusters than other fields (like  ALH-4 and ALH-8). Finally, we notice filament-like structures, i.e. clustering of groups at similar redshift ranges. For instance, ALH-3 clearly shows an over density of groups at $z\sim 0.8$, ALH-8 at $z\sim 0.5$ and $z\sim 0.65$, etc.

\begin{figure}
\centering
\includegraphics[clip,angle=0,width=1.0\hsize]{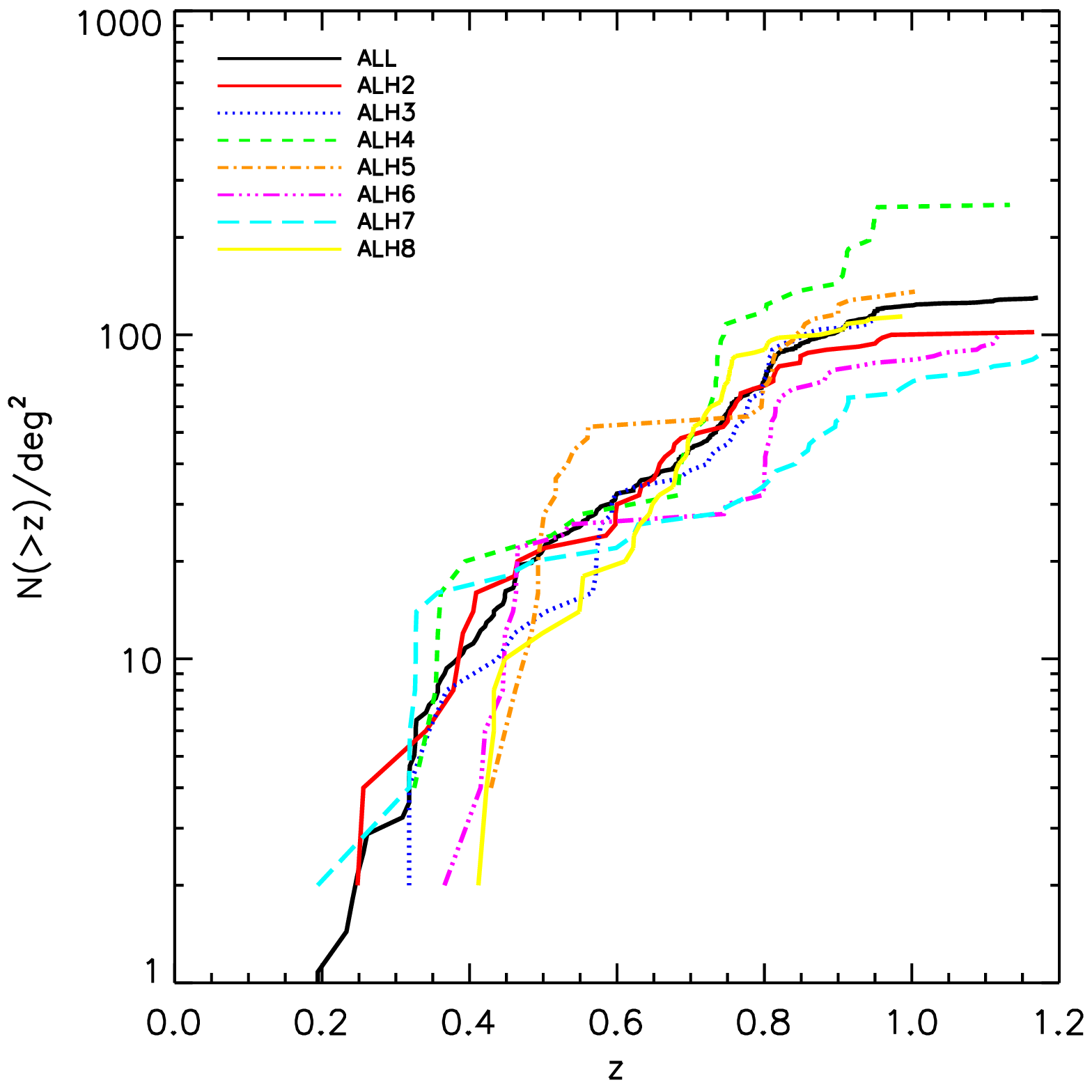} 
\caption{Cumulative number of clusters per square degree as a function of redshift for the whole sample (black solid line) and each separated field (ALH-2, red solid line; ALH-3, blue dotted line; ALH-4, green dashed line; ALH-5, dotted-dashed orange line; ALH-6, three dotted dashed magenta line; ALH-7,  long dashed cyan line; ALH-8, solid yellow line). The changes of the slope indicate the presence of clustering at a particular redshift range, making evident the presence of the cosmic variance. It is noticeable the departure of the ALH-4 field with respect to the other ones.}
\label{fig:zdist}
\end{figure}

In an effort to inspect these features in more detail, we have plotted in Fig. \ref{fig:zdist} the cumulative number of clusters per square degree as a function of redshift  for the whole sample (black line) and each individual field. We confirm the previous analysis. The presence of substructure in the COSMOS field (ALH-4) is evident, finding three main sharp increases in the cumulative function at $\sim$ 0.4, 0.7 and 0.9. We notice several changes of the slope at different redshifts for different fields. In particular, in ALH-7, we distinguish a smooth increase in the slope of the cumulative function between $\sim$ 0.55 and 0.8, already predicted by \cite{arnalte-mur14}.

\subsubsection{Presence and absence of a red sequence}

The existence of a tight red sequence (RS) in galaxy clusters has been widely studied in numerous works (e.g. \citealt{lopez-cruz04,mei06,ascaso08}, and references herein). The main theories of formation of galaxies claim that elliptical galaxies are formed at high redshifts in the most massive environments and evolve passively since then.  One of the safest probes on the `universal' existence of the RS down to $M>10^{14}\rm M_{\odot}$ and up to moderate redshifts ($z<1.6$) is that clusters detected with methods non-dependent on the RS, such as the X-rays or SZ, still display a well-defined RS.  However, there are a number of observational works that have demonstrated the change of paradigm at high ($z >1$) redshift \citep{brodwin13,mei14}, and lower masses ($\sim$ several times $10^{13}\rm M_{\odot}$, \citealt{finoguenov07}). Star formation increases in these almost unexplored ranges of mass and redshift and, consequently, the galactic population of these structures is significantly bluer and of later-type.

We have examined the nature of the RS in the detections found in the ALHAMBRA survey. The galaxies attributed to each cluster candidate have been selected as those galaxies within 1 Mpc radial distance from the clusters centre at the redshift of the cluster, and accomplishing the condition set in Eq. \ref{eq:zdiff}. Also, in order to make a clean selection, we have considered a galaxy to belong to the cluster if \emph{odds}$>0.5/(1+z)$ and $Stellar\_Flag<0.7$.

\begin{figure}
\centering
\includegraphics[clip,angle=0,width=1.\hsize]{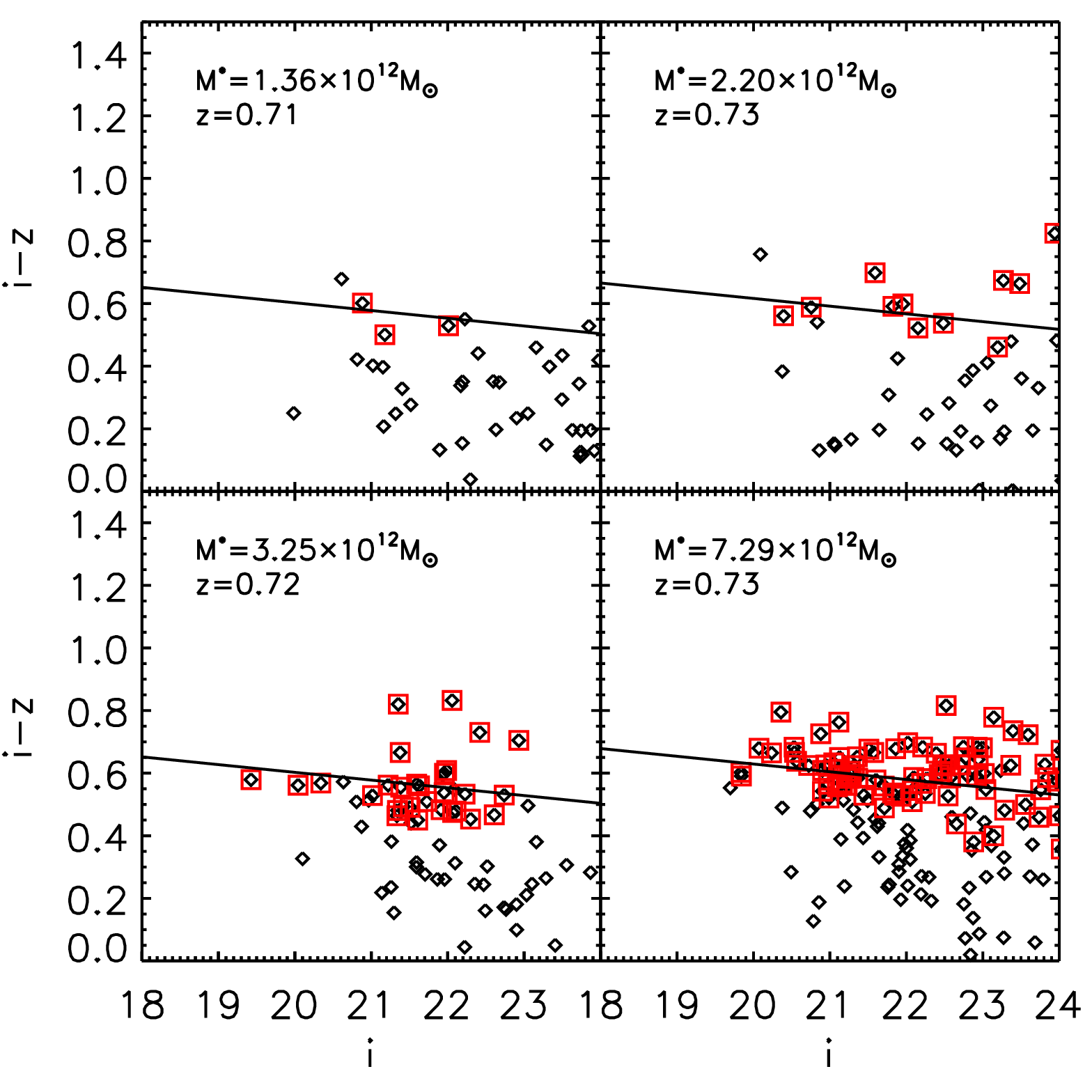} 
\caption{Colour-magnitude relation of four of the detections obtained in the ALHAMBRA belonging to four different bins from the smallest (top left) to the richest (bottom right). Only galaxies with  \emph{odds}$>0.5/(1+z)$ are shown. The red squares refer to those with elliptical ($t_b<5$) spectral type according to the BPZ2.0 classification. The solid lines indicates the expected colour-magnitude relation at the redshift of the cluster (section \S3). The fraction of blue galaxies appears to be higher for the less massive structures and few early-type galaxies are already in place on the red sequence in the same clusters.}
\label{fig:cmr}
\end{figure}

In Fig. \ref{fig:cmr}, we show the RS of four different detections belonging to different increasing mass bins (from less massive in the top left panel to more massive clusters in the bottom right panel) at a similar redshift ($\sim$ 0.7). The red squares refer to those galaxies classified as early-type (spectral template, $t_b$, $<5$) by BPZ2.0. Looking at these examples, it becomes clear the difference between the least and most massive structures. While the richer clusters exhibit a well-formed red sequence, the less massive clusters display a few red galaxies in place. 

In order to quantify this, we have measured the \emph{photometric} blue fraction for each cluster at a given redshift. This fraction has been defined as:

\begin{equation}
f_B=\frac{N_B}{N_T}
\end{equation}
\noindent where $N_B$ refer to the blue galaxies, considered as those with a bluer colour than the main expected colour of the red sequence at its redshift minus a typical dispersion. For clusters at redshift $<$0.7, we will consider the $g-i$ colour, while for clusters $>$0.7, we will use the $i-z$ colour.  $N_T$ refers to the total number of galaxies considered to belong to the cluster. 

This quantity has been measured down to a fixed absolute magnitude, $M_i=-19.6$ which corresponds to $i\sim 24.5$, the magnitude limit of ALHAMBRA at redshift $\sim1.0$, to be able to account for all the galaxies independent of the redshift dimming. Also, no \emph{odds} cut has been performed here. 

\begin{figure}
\centering
\includegraphics[clip,angle=0,width=1.\hsize]{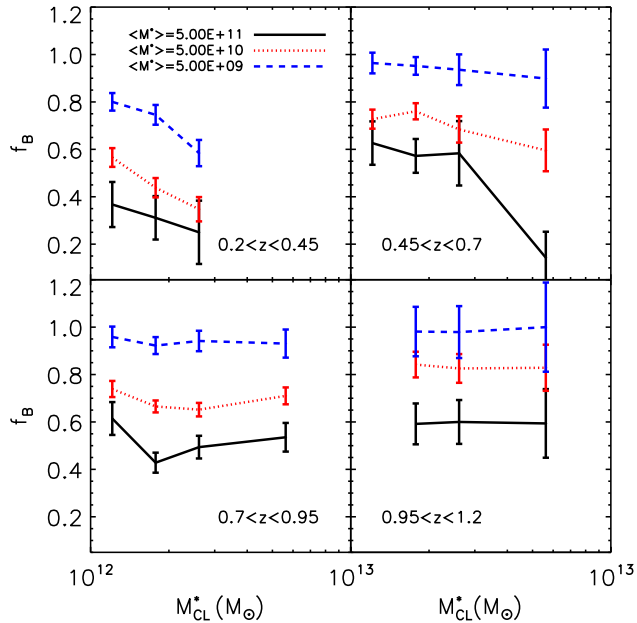} 
\caption{Relationship between the total cluster stellar mass and the   \emph{photometric} blue fraction for the ALHAMBRA cluster and group sample stacked into different redshift  bins. The different lines refer to different galaxy masses bins ($1\times 10^{9} <M^*<1 \times 10^{10}$, dashed blue line; $1\times 10^{10} <M^*<1 \times 10^{11}$, dotted red line;$1\times 10^{11} <M^*<1 \times 10^{12}$, solid black line). The galaxy member candidates have been selected within 1 Mpc distance of the center and performing the photometric redshift cut specified in Eq. \ref{eq:zdiff}. No \emph{odds} has been performed. Blue galaxies are considered those with a bluer colour than the main expected colour at its redshift minus 0.3 magnitudes. The error bars include simple Poissonian statistics and field contamination errors.}
\label{fig:SMfb}
\end{figure}

We have also investigated the dependence of the \emph{photometric} blue fraction with the cluster/group environment, as a function of the redshift. To do this, we have stacked galaxy clusters and groups in bins of redshift and we have computed $f_B$ as a function of cluster total stellar mass, for different galaxy masses ranges, which is shown in Fig \ref{fig:SMfb}. 

We expect the possible contamination of the non-member galaxies to be almost negligible, due to the excellent accuracy of the photometric redshifts of the survey, particularly for the bright end of the cluster distribution. In order to quantify this, we have  computed the blue fraction of all the clusters in the mock catalogue given the spectroscopic redshift and photometric redshifts separately and we have estimated the contamination as the absolute difference of their mean values for different redshift bins. As expected, the field contamination source of error is more than 10 times smaller than the poissonian errors. Both source of errors have been included in the error bars in Fig \ref{fig:SMfb}.

As expected,  the mean $f_B$ increases as a function of redshift both in clusters and groups for a fixed galaxy mass. Also, despite the small size of the sample, we observe a tendency between the \emph{photometric} blue  fraction and the cluster/group total stellar mass for a fixed galaxy mass. We observe, with more than 2$\sigma$ confidence, that the low-mass groups have higher $f_B$ compared to more massive clusters,  at least up to redshift 0.45. 

Interestingly, the slope of the dependence of the blue fraction with the cluster richness seems also to evolve with redshift. At high redshift, the slope seems to be flat, whereas it becomes steeper at lower redshift. This result would be in agreement with other observational results that have suggested two different regimes for galaxy evolution, being the environment only active at low redshift \citep{peng10}. 

While this is a tentative piece of evidence, the small size of the sample does not allow to throw a more definitive conclusion. Future surveys, such as the J-PAS, will provide hundreds of thousands of clusters and groups with even better photometric redshift precisions, increasing the statistics and diminishing the error bars. As a result, it will confirm with high confidence these tendencies.

The aim of this work is to describe the cluster and group sample detected in the ALHAMBRA survey and we do not attempt to explore this outcome in detail here. A more quantitative analysis will be carried out in a separate paper (D\'iaz-Garc\'ia et al. in prep).

\section{Summary and conclusions}

In this work, we present the first release of the optical galaxy clusters and groups catalogue in the ALHAMBRA survey. According to simulations, we are able to sample the mass function down to $\sim 6\times10^{13}\rm M_{\odot}$, up to redshift 0.8, with both completeness and purity higher than 80\% and down to $\sim 3\times10^{13}\rm M_{\odot}$, with detection rates $>$70\% to the same redshift. At higher redshift, the mass threshold increases, being able to detect structures with masses $> 1\times10^{14}\rm M_{\odot}$ up to redshift 1 in the first case, and $> 6\times10^{13}\rm M_{\odot}$ up to redshift $\sim 1$ and $> 1\times10^{14}\rm M_{\odot}$ up to redshift $\sim $1.2, in the second case. Additionally, we have compared our detections with other cluster samples obtained from a variety of datasets and techniques, achieving a very good agreement when matched to our sample, confirming the completeness of our results. Additionally, we have shown how the detections that are not found in other work are mainly located at high redshift and low masses, confirming the reliability of the results.

This mass limit threshold conveys an important result, which is the evidence that deep multi-band medium and narrow-band filters allow us to sample the mass function with reliability down to smaller masses than deep broad-band surveys.  For instance, we detected galaxy clusters and groups in two optical broad-band surveys: the DLS  \citep{ascaso14a} and the CARS  \citep{ascaso12} surveys with the same methodology, the BCF. We were only able to obtain mass limits of $> 1.2\times10^{14}\rm M_{\odot}$  and $> 4\times10^{14}\rm M_{\odot}$ with completeness and purity rates $>$70\% and $>$80\% respectively, even if the data was $\sim 2.5$ magnitudes deeper for the DLS and similar depth for the CARS.

A second important consequence is the fact that the high photometric redshift accuracy that such surveys provide, allows us to obtain a reliable determination of the membership of the cluster. This directly translates into one of the highest accuracies at calibrating the redshift and mass of the clusters reachable with optical data up to date. In this work, we have been able to calibrate the total stellar mass - dark matter halo relation with a precision of $\sigma_{M_h | M^*_{CL}}\sim 0.25-0.35$ dex down to $\sim 3\times10^{13}\rm M_{\odot}$, which is very similar to what other techniques (optical, X-rays, SZ) have found for, at least, two orders of magnitude higher mass limit. Moreover, we have measured the dispersion at recovering the overall redshift of the cluster or group, obtaining a main dispersion of $\sigma_{\rm NMAD} \sim$0.006.

Since few optical surveys have been able to sample the mass range to these lower limits, the evolution of properties of the galaxies in those groups has not been widely explored due to, on one hand, the inability to obtain a complete sample and, on the other, due to a possible bias of selecting `red clusters' when using other techniques related with the CMR. In this study, we have preliminarily looked into the overall characteristics of the galactic population of these new set of groups detected in the ALHAMBRA survey, reaching very low mass limit thresholds.

In particular, we have used the optical ALHAMBRA group sample to report the visible increase of the fraction of blue galaxies in low-mass groups. Indeed, we find a significantly lower fraction of blue galaxies in $z<0.5$  ALHAMBRA groups consistent with more efficient environmental quenching in the local Universe \citep{peng10} and in agreement with other results on the blue fraction in more massive systems \citep{raichoor12,raichoor14}.

These results become paramount for future applications of this kind of survey. In particular, the J-PAS survey \citep{benitez14} is a survey which will be starting in 2015 and will image 8600 square degrees with 54 optical narrow-bands, down to $r\sim 23.5$. The expected photometric redshift accuracy of this survey is 0.003 and therefore, the expected number of groups and clusters that we can detect with reliability amounts to higher numbers than other similar projects aiming to go deeper with fewer, broader bands (Ascaso et al. in prep). As a result, this kind of data will also allow to confirm with high significance ($>$10$\sigma$) the possible galaxy evolutionary mechanisms happening in clusters and groups.

Finally, as other works have already claimed \citep{lopez-sanjuan14,arnalte-mur14,molino14}, the study of the distribution of the cluster properties in each of the different fields of ALHAMBRA sets evidence on the cosmic variance at the level of the clustering of clusters. We notice striking differences between different fields in terms of mass and redshift distribution. In future work, we will explote both the large scale properties of these structures and the properties of the galactic population of these detections

\section*{Acknowledgements}
This work is based on observations collected at the Centro Astron—mico Hispano Alem‡n (CAHA) at Calar Alto, operated jointly by the Max-Planck Institut fŸr Astronomie and the Instituto de Astrof'sica de Andaluc'a (CSIC). We thank the anonymous referee for his/her valuable comments that help to improve this paper. BA thanks Carlton Baugh for his useful comments on an earlier version of the draft. We acknowledge support from the Spanish Ministry for Economy and Competitiveness and FEDER funds through grants AYA2010-22111-C03-02, AYA2010-15169, AYA2012-30789, AYA2013-48623-C2-2, AYA2013-42227-P, AYA2013-40611-P, AYA2011-29517-C03-01, AYA2014-58861-C3-1, AYA2010-15081, Generalitat Valenciana projects PROMETEOII/2014/060, Junta de Andaluc\'{\i}a grant TIC114, JA2828, Arag\'on Gorvernment - Research Group E103. MP acknowledges financial support from JAE-Doc program of the Spanish National Research Council (CSIC), co-funded by the European Social Fund. PAM acknowledges support from ERC StG Grant DEGAS-259586 and from the Science and Technology Facilities Council grants ST/K003305/1 and ST/L00075X/1. This work used the DiRAC Data Centric system at Durham University, operated by the Institute for Computational Cosmology on behalf of the STFC DiRAC HPC Facility (www.dirac.ac.uk). This equipment was funded by BIS National E-infrastructure capital grant ST/K00042X/1, STFC capital grant ST/H008519/1, and STFC DiRAC Operations grant ST/K003267/1 and Durham University. DiRAC is part of the National E-Infrastructure. BA dedicates this paper to the memory of Javier Gorosabel, whose wisdom and sense of humour will always be missed.

\section*{Affiliations}

$^{1}$GEPI, Observatoire de Paris, CNRS, Universit\'e Paris Diderot, 61, Avenue de l'Observatoire 75014, Paris  France\\
$^{2}$Instituto de Astrof\'isica de Andaluc\'ia (IAA-CSIC), Glorieta de la Astronom\'ia s/n, 18008, Granada, Spain\\
$^{3}$Instituto de F\'{\i}sica de Cantabria (CSIC-UC), Avenida de los Castros s/n, E-39005 Santander, Spain\\
$^{4}$Unidad Asociada Observatori Astron\`omic (UV-IFCA), C/ Catedr\'atico Jos\'e Beltr\'an 2, E-46980 Paterna, Spain\\
$^{5}$Observatori Astron\`omic de la Universitat de Val\`encia,  C/ Catedr\'atico Jos\'e Beltr\'an 2, E-46980 Paterna, Spain\\
$^{6}$Institute for Computational Cosmology, Department of Physics, Durham University, South Road, Durham DH1 3LE, UK\\
$^{7}$Centro de Estudios de F\'isica del Cosmos de Arag\'on, Plaza San Juan 1, 44001 Teruel, Spain\\
$^{8}$Instituto de Astronomia, Geof\'isica e Ci\^encias Atmosf\'ericas, Universidade de S\~ao Paulo, Cidade Universit\'aria, 05508-090, S\~ao Paulo, Brazil\\
$^{9}$Observat\'orio Nacional-MCT, Rua Jos\'e Cristino, 77. CEP 20921-400, Rio de Janeiro-RJ, Brazil\\
$^{10}$Department of Physics and Astronomy, University College London, Gower Street, London WC1E 6BT, UK\\
$^{11}$University Denis Diderot, 4 rue Thomas Mann, 75205 Paris, France\\
$^{12}$Departament dÕAstronomia i Astrof\'{\i}sica, Universitat de Val\`encia, E-46100, Burjassot, Spain\\
$^{13}$Dept. of Astronomy, University of Michigan, Ann Arbor, MI 48109\\
$^{14}$Eureka Scientific, Oakland, California, 94602-3017\\
$^{15}$Instituto de Astrofisica de Canarias, Via Lactea s/n, La Laguna, 38200 Tenerife, Spain\\
$^{16}$Departamento de Astrofõsica, Facultad de F«õsica, Universidad de la Laguna, 38200 La Laguna, Spain\\
$^{17}$Department of Theoretical Physics, University of the Basque Country UPV/EHU, 48080 Bilbao, Spain\\
$^{18}$IKERBASQUE, Basque Foundation for Science, Bilbao, Spain\\
$^{19}$Departamento de F\'isica At\'omica, Molecular y Nuclear, Facultad de F\'isica, Universidad de Sevilla, 41012 Sevilla, Spain\\
$^{20}$Institut de Ci\`encies de l'Espai (IEEC-CSIC), Facultat de Ci\`encies,  Campus UAB, 08193 Bellaterra, Spain\\
$^{21}$Departamento de Astronom\'ia, Pontificia Universidad Cat\'olica. 782-0436 Santiago, Chile\\
$^{22}$Instituto de F\'{\i}sica Te\'orica, (UAM/CSIC), Universidad Aut\'onoma de Madrid, Cantoblanco, E-28049 Madrid, Spain \\
$^{23}$Campus of International Excellence UAM+CSIC, Cantoblanco, E-28049 Madrid, Spain

\end{document}